\documentclass[11pt,a4paper]{article}
\usepackage{graphicx} 
\usepackage{amsfonts,amsmath,amssymb} 
\usepackage{float} 
\usepackage[margin=2cm]{geometry} 
\usepackage{longtable} 
\usepackage{color,url}
\usepackage{colortbl}
\usepackage{multirow}
\usepackage{verbatim}
\usepackage[utf8]{inputenc}
\usepackage[colorlinks=true
,urlcolor=blue
,anchorcolor=blue
,citecolor=blue
,filecolor=blue
,linkcolor=blue
,menucolor=blue
,linktocpage=true
,pdfproducer=medialab
,pdfa=true
]{hyperref}
\usepackage{cite}

\usepackage{epsfig,psfrag,rotating,soul}
\usepackage{rotfloat}
\usepackage{epsf}
\usepackage{graphics}
\usepackage{xcolor}
\usepackage{graphicx}

\begin{document}
\thispagestyle{empty}

\vspace*{2.5cm}

\vspace{0.5cm}

\begin{center}

\begin{Large}
\textbf{\textsc{Flavour Anomalies: A comparative analysis using a machine learning algorithm}}
\end{Large}

\vspace{1cm}

{\sc
S.~Pe{\~n}aranda$^{a,b}$%
\footnote{\tt \href{mailto:siannah@unizar.es}{siannah@unizar.es}}%
J. Alda$^{b, c, d}$%
\footnote{\tt \href{mailto:jorge.alda@pd.infn.it}{jorge.alda@pd.infn.it}}%
, A.~Mir$^{a,b}$%
\footnote{\tt \href{mailto:amir@unizar.es}{amir@unizar.es}}%
}

\vspace*{.7cm}

{\sl
$^a$Departamento de F{\'\i}sica Te{\'o}rica, Facultad de Ciencias,\\
Universidad de Zaragoza, Pedro Cerbuna 12,  E-50009 Zaragoza, Spain

\vspace*{0.1cm}

$^b$Centro de Astropart{\'\i}culas y F{\'\i}sica de Altas Energ{\'\i}as (CAPA), \\
Universidad de Zaragoza, Zaragoza, Spain

\vspace*{0.1cm}

$^c$Dipartimento di Fisica ed Astronomia ``Galileo Galilei'',\\
Università degli Studi di Padova, via F. Marzolo 8, 35131 Padova, Italy

\vspace*{0.1cm}

$^d$Istituto Nazionale di Fisica Nucleare (INFN), Sezione Padova,\\
via F. Marzolo 8, 35131 Padova, Italy

}

\end{center}

\vspace*{0.1cm}

\begin{abstract}
\noindent

We present an analysis on flavour anomalies in semileptonic
  rare $B$-meson decays using an effective field theory approach and
  assuming that new physics affects only one generation in the
  interaction basis and non-universal mixing effects are generated by
  the rotation to the mass basis. A global fit to experimental data is
  performed, focusing on LFU ratios $R_{D^{(*)}}$ and $R_{J/\psi}$ and
  branching ratios that exhibit tensions with Standard Model predictions
  on $B \rightarrow K^{(*)} \nu \bar{\nu}$ decays. In our analysis, we use a Machine
  Learning Montecarlo algorithm, a framework that
  emulates the highly non-Gaussian structure of the likelihood landscape
  with minimal training cost. This method enables the generation of
  high-resolution confidence regions and detailed correlation analyses.
  By comparing three different scenarios, we show that the one that introduces only mixing
  between the second and third quark generations and no mixing in the
  lepton sector, as well as independent coefficients for the singlet and
  triplet four fermion effective operators, provides the best fit to the 
  experimental data. A comparison with previous results is performed. We
  highlight the key strengths of the Machine Learning framework in our analysis. 

\end{abstract}

\def\thefootnote{\arabic{footnote}}
\setcounter{page}{0}
\setcounter{footnote}{0}

\newpage

\section{Introduction}
\label{intro}

In the last decade several experimental anomalies related to flavour physics and, in particular, in $B$-meson decays, have been observed~\cite{LHCb:2014cxe,LHCb:2014vgu,LHCb:2015wdu,LHCb:2015svh,LHCb:2016ykl,CMS:2014xfa,LHCb:2015gmp,LHCb:2016due,LHCb:2017rmj,LHCb:2017myy,LHCb:2017avl,LHCb:2017smo,LHCb:2018rym,LHCb:2019hip,CMS:2023vgr,Belle-II:2024ami,LHCb:2024jll,CMS:2024syx}. The disagreement between the theoretical predictions in the Standard Model (SM) and the experimental measurements is a clear open window for searches of physics beyond the SM. A model-independent analysis of the contributions to $B$-physics observables displaying those discrepancies is highly recommended. The Effective Field Theory (EFT) approach allows us to perform this kind of analysis by focusing on the relevant degrees of freedom and describing the physics one is interested in~\cite{Alda:2021rgt,Alda:2021ruz,Alda:2020okk,AldaGallo:2021cku}. In this way, we can study the allowed regions of parameter space which are compatible with experimental results.

In order to better identify the possible deviations from the SM behaviour, one can construct ratios of branching ratios, which are ``cleaner'' observables as hadronic uncertainties largely cancel. For example, the $\ensuremath{R_{D^{(*)}}}$ ratios, 
\begin{equation}
\ensuremath{R_{D^{(*)}}}^\ell = \frac{\mathrm{BR}(B \to D^{(*)} \tau \bar{\nu}_\tau ) }{[\mathrm{BR}(B \to D^{(*)} e \bar{\nu}_e) + \mathrm{BR}(B \to D^{(*)} \mu \bar{\nu}_\mu)]/2}\,,
\end{equation}
whose values in the SM are~\cite{Jaiswal:2017rve}:
 \begin{equation}
     R_D^{SM}= 0.299\pm0.004\,, \quad \quad R_{D^*}^{SM}= 0.257\pm0.005 \,.
 \end{equation}

The latest measurements of the $R_{D^{(*)}}$ ratios have been performed at Belle II~\cite{Belle-II:2024ami},
\begin{equation}
    R_{D^*}^\ell = 0.262^{+0.041}_{-0.039}{}^{+0.035}_{-0.032}\,,
\end{equation}
and LHCb~\cite{LHCb:2024jll},
\begin{align}
    R_D^\ell &= 0.249 \pm 0.043 \pm 0.047\,, \nonumber\\
    R_{D^*}^\ell &= 0.402 \pm 0.081 \pm 0.085\,,
\end{align}
bringing the world average to~\cite{hflav24}
\begin{align}
    R_D^\ell &= 0.342 \pm 0.026 \,, \nonumber\\
    R_{D^*}^\ell &= 0.287 \pm 0.012 \,,
\end{align}
which are $1.6\,\sigma$ and $2.5\,\sigma$ above the SM, respectively, with a combined tension of $3.31\,\sigma$.

Another class of $B$ meson observables is $R_{J/\psi}$, defined as the Lepton Flavour Universality (LFU) ratio,
\begin{equation}
    R_{J/\psi} = \frac{\mathrm{BR}(B_c\to J/\psi \,\tau \,\nu_\tau)}{\mathrm{BR}(B_c\to J/\psi \,\mu \,\nu_\mu)}\,.
\end{equation}
At the quark level, it corresponds to a $b\to c \ell \nu$ transition, which is the same transition as the $R_D$ and $R_{D^*}$ ratios. As such, it is natural to examine the $R_{J/\psi}$ ratio looking for a similar anomaly. Indeed, the SM prediction is~\cite{Tang:2022nqm}
\begin{equation}
    R_{J/\psi}^\mathrm{SM} = 0.258 \pm 0.004\,,
\end{equation}
while the experimental measurement at LHCb~\cite{LHCb:2017vlu}
\begin{equation}
    R_{J/\psi}^\mathrm{LHCb} = 0.71 \pm 0.17 \pm 0.18\,,
\end{equation}
displaying a $1.8\,\sigma$ tension. It is interesting to note that the experiments detect an excess of tauonic decays (or equivalently a defect of light lepton decays), which is also the case for the $R_D$ and $R_{D^*}$ anomalies. However, this tension is not present in the recent measurement at CMS~\cite{CMS:2023vgr}
\begin{equation}
    R_{J/\psi}^\mathrm{CMS} = 0.17^{+0.18}_{-0.17}{}^{+0.21}_{-0.22}{}^{+0.19}_{-0.18}\,,
\end{equation}
which is only $0.3\,\sigma$ away from the SM prediction. The na\"ive average of both measurements results in~\cite{Iguro:2024hyk}
\begin{equation}
    R_{J/\psi} = 0.52 \pm 0.20\,,
    \label{eq:averageRJpsi}
\end{equation}
consistent with the SM prediction at the $1.3\,\sigma$ level.

Besides, Belle II has also reported an excess in the $B^+\to K^+\nu\bar{\nu}$ decay~\cite{Belle-II:2023esi}, combining inclusive and hadronic tagging,
\begin{equation}
    \mathrm{BR}(B^+\to K^+\nu \bar{\nu}) = (2.3\pm0.7)\times10^{-5}\,,
\end{equation}
which present a $2.7\,\sigma$ excess as compared to the SM prediction~\cite{Bause:2023mfe},
\begin{equation}
    \mathrm{BR}(B^+\to K^+\nu\bar{\nu})_\mathrm{SM} = (0.429 \pm 0.013 \pm 0.02)\times 10^{-5}\,.
\end{equation}

For the decay into a vector kaon, the $90\%$ confidence level limit set by Belle~\cite{Belle:2017oht},
\begin{equation}
    \mathrm{BR}(B^0 \to K^{*0}\nu\bar{\nu}) < 1.8\times 10^{-5}\,,\qquad  \mathrm{BR}(B^+ \to K^{*+}\nu\bar{\nu}) < 6.1\times 10^{-5}\,,
\end{equation}
is at least still a factor of 2 away from the SM prediction~\cite{Bause:2021cna},
\begin{equation}
    \mathrm{BR}(B \to K^{*}\nu\bar{\nu})_\mathrm{SM} = (0.824\pm0.099)\times10^{-5}\,.
\end{equation}

The experimental data has been used in recent years to constrain New
Physics (NP) models. Several global fits have been performed in the
literature~\cite{Calibbi:2015kma,Capdevila:2017bsm,Celis:2017doq,Alok:2017sui,Camargo-Molina:2018cwu,hflav24,Datta:2019zca,Aebischer:2019mlg,Aoude:2020dwv,Alda:2020okk,Alda:2021rgt,Alda:2021ruz,AldaGallo:2021cku,Tang:2022nqm,Alguero:2022wkd,SinghChundawat:2022zdf,Grunwald:2023nli,Bartocci:2023nvp,Elmer:2023wtr,Capdevila:2023yhq,Grunwald:2024yuq,Iguro:2024hyk},
and references therein. The present work includes in the global analysis
the recent experimental measurements of $R_{D^{*}}$~\cite{hflav24} and
$R_{J/\psi}$~\cite{Iguro:2024hyk}, which were not included in~\cite{Alda:2021rgt}. Besides, the latest results for $B^+\to K^+\nu\bar{\nu}$ decay are also included in the discussion. Crucially, the experimental landscape has undergone a fundamental shift since our previous work: the updated $R_K$ and $R_{K^*}$ measurements from 2022~\cite{LHCb:2022qnv} have resolved the long-standing anomalies in $b\to s\ell^+\ell^-$ transitions, bringing these ratios into agreement with SM predictions. This development, combined with the emergence of a significant excess in $B^+\to K^+\nu\bar{\nu}$, motivates a complete reconsideration of the operator structure. While the same SMEFT operators that were previously favored remain relevant, their relative importance and the optimal flavor structure have changed dramatically.

In this paper, we demonstrate that this new experimental situation requires allowing independent Wilson coefficients for singlet and triplet operators ($C_1 \neq C_3$), a key distinction from our previous scenarios where we assumed $C_1 = C_3$. This generalization, which we implement in our new Scenario III, provides substantially better fits to the data and reveals a qualitatively different phenomenology. Specifically, it allows us to simultaneously describe the persistent anomalies in $b\to c\tau\nu$ and the new excess in $b\to s\nu\bar{\nu}$ channels while maintaining consistency with the now SM-like $b\to s\,\ell^+\ell^-$ observables.

The paper is organized as follows. Section~\ref{sec:EFT} provides a summary of the relevant terms of the Weak Effective Theory for our analysis, as well as few details on how to perform the statistical fit of the Wilson coefficients involved. We analyse the $b\to c \tau \bar{\nu}$ and the $B\to K^{(*)} \nu \bar{\nu}$ decays, providing the best fit predictions for $R_{D^{*}}$ and $R_{J/\psi}$ observables and for the $B^+\to K^+\nu\bar{\nu}$ and $B^0\to K^{*0}\nu\bar{\nu}$ branching ratios. Section~\ref{sec:GF} is devoted to the global fits, by including some details of the Machine Learning Montecarlo algorithm used up. At the end of this section we discuss the correlation for the $R_{D^*}$ observable and $B^+\to K^+ \nu\bar{\nu}$ decay. Finally, conclusions are presented in section~\ref{sec:CONCLU}.

\section{Effective field theories}
\label{sec:EFT}

The Weak Effective Theory (WET) is formulated at an energy scale below the electroweak scale and the heavy SM states ($t$, $h$, $W^\pm$ and $Z^0$) are integrated out. The NP resides at an energy that is greater than the experimentally accessible energies, and the Wilson coefficients are a measure of the strength with which the NP couples to ordinary particles. The dimension six operators are constructed from light SM fields. In this section we only present the relevant terms for our analysis from the WET Lagrangian~\cite{Buras:1998raa,Aebischer:2015fzz,Aebischer:2017gaw}. We start by analyzing the $b\to c \tau \bar{\nu}$ decays and then discuss the $B\to K^{(*)} \nu \bar{\nu}$ decays. The best fit predictions for $R_{D^{*}}$ and $R_{J/\psi}$ observables and for the $B^+\to K^+\nu\bar{\nu}$ and $B^0\to K^{*0}\nu\bar{\nu}$ branching ratios are obtained.

\subsection{\texorpdfstring{\boldmath{$b\to c \tau \bar{\nu}$}}{b c tau nu} decays}
\label{sec:eft_rd}

In the WET below the electroweak scale, the expressions for $R_{J/\psi}$ were given by~\cite{Harrison:2020gvo,Tang:2022nqm}, and for $R_{D^{(*)}}$ by~\cite{Tanaka:2012nw}. The relevant effective Lagrangian is
\begin{equation}
    \mathcal{L}_{b\to c\tau\nu} = -\frac{4G_F}{\sqrt{2}}V_{cb}\left[(1+C_{VL}) \,O_{VL} + C_{VR}\,O_{VR} + C_{SL}\,O_{SL} + C_{SR}\,O_{SR} + C_T \,O_T \right]+\mathrm{h.c.}\,,
    \label{eq:Lag_bctaunu}
\end{equation}
where $G_F$ is the Fermi constant, $V_{cb}$ is an element of the CKM matrix and the effective operators are defined as
\begin{align}
    O_{VL} &= (\bar{c}_L\gamma^\mu b_L)(\bar{\tau}_L\gamma_\mu \nu_L)\,, &\qquad\qquad O_{VR} &= (\bar{c}_R\gamma^\mu b_R)(\bar{\tau}_L\gamma_\mu \nu_L)\,, \nonumber\\
    O_{SL} & = (\bar{c}_R b_L)(\bar{\tau}_R \nu_L)\,, &\qquad\qquad O_{SR} & = (\bar{c}_L b_R)(\bar{\tau}_R \nu_L)\,, \nonumber\\
    O_T &= (\bar{c}_R \sigma^{\mu\nu} b_L)(\bar{\tau}_R \sigma_{\mu\nu} \nu_L)\,,
\end{align}
being their corresponding Wilson coefficients $C_{VL}, C_{VR}, C_{SL}, C_{SR}$ and $C_T $, respectively.

In the SM, the $b\to c\ell\nu$ transitions at tree level are mediated by the exchange of a $W$ boson, which in the WET is integrated out contributing to the $C_{VL}^\mathrm{tot} = 1 + C_{VL}$ Wilson coefficient. This contribution has already been separated in Lagrangian~\eqref{eq:Lag_bctaunu}, and therefore any nonzero value of the Wilson coefficients in the Lagrangian will be due only to NP. The dominant contribution will come from the interference term between $C_{VL}$ and the SM. Furthermore, the scalar contributions $C_{SL}$ and $C_{SR}$ are helicity-suppressed by the mass of the lepton~\cite{Sakaki:2013bfa}.

The NP contributions at an energy scale $\Lambda \sim \mathcal{O}(\mathrm{TeV})$ are defined via the Standard Model Effective Field Theory (SMEFT) Lagrangian~\cite{Grzadkowski:2010es}. The relevant terms for our analysis can be written as,
\begin{equation}
\mathcal{L}_\mathrm{SMEFT} = \frac{1}{\Lambda^2}\left(\ensuremath{C_{\ell q(1)}^{ijkl}}\, O_{\ell q(1)}^{ijkl} + \ensuremath{C_{\ell q(3)}^{ijkl}}\,  O^{ijkl}_{\ell q(3)} \right)  \ ,
\label{eq:Lag_SMEFT_GF}
\end{equation}
where $\Lambda$ is the energy scale, $\ell$ and $q$ are the lepton and quark $SU(2)_L$ doublets, ${i,j,k,l}$ denote generation indices, and the dimension six operators are defined as 
\begin{equation}
O_{\ell q(1)}^{ijkl} = (\bar{\ell}_i \gamma_\mu \ell_j)(\bar{q}_k \gamma^\mu  q_l),\qquad\qquad O_{\ell q(3)}^{ijkl}= (\bar{\ell}_i \gamma_\mu \tau^I \ell_j)(\bar{q}_k \gamma^\mu \tau^I q_l) ,
\label{eq:operdim6}
\end{equation}
with $\tau^I$ being the Pauli matrices and $C_{\ell q(1)}^{ijkl}$ and
$C_{\ell q(3)}^{ijkl}$ are the corresponding Wilson coefficients. Here
we only include the SMEFT operators containing two left-handed quarks
and two left-handed leptons motivated by our interest in the $B$ anomalies. 

The tree-level matching between the WET and SMEFT coefficients at the
scale $\mu=M_Z$ is given by~\cite{Jenkins:2017jig}: 
\begin{align}
    C_{VL} &= -\frac{\sqrt{2}}{4G_F \Lambda^2 V_{cb}}\left(2 C_{\ell q(3)}^{3323} - 2C_{\varphi q(3)}^{23} \right)\,, \qquad &&C_{VR} = \frac{\sqrt{2}}{4G_F \Lambda^2 V_{cb}} C_{\varphi u d}^{23}\nonumber\\
    C_{SL} &= -\frac{\sqrt{2}}{4G_F \Lambda^2 V_{cb}} C_{\ell e qu(1)}^{3332\,*}\,,\qquad
    &&C_{SR} = -\frac{\sqrt{2}}{4G_F \Lambda^2 V_{cb}}  C_{ledq}^{3332\,*}\,,\nonumber\\
    C_T &= -\frac{\sqrt{2}}{4G_F \Lambda^2 V_{cb}} C_{\ell e qu(3)}^{3332\,*}\,.
    \label{eq:matching}
\end{align}
It is important to note that the contributions from scalar-quark operators $C_{\varphi q(3)}$ and $C_{\varphi u d}$ are universal in the lepton flavour, and therefore can not contribute to the $B$-meson anomalies. The contributions from $C_{\ell equ(1)}$, $C_{\ell equ(3)}$ and $C_{\ell edq}$ involve right-handed fermions, which are suppressed because they do not interfere with the SM electroweak interactions.

\begin{figure}
    \centering
    \begin{tabular}{ccc}
     \includegraphics[width=0.3\textwidth]{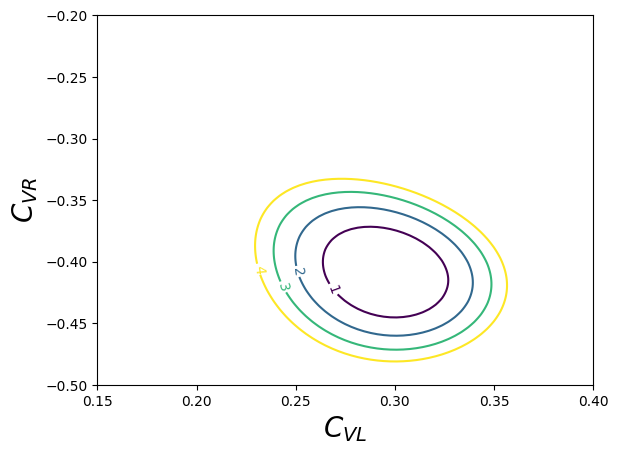}&
     \includegraphics[width=0.3\textwidth]{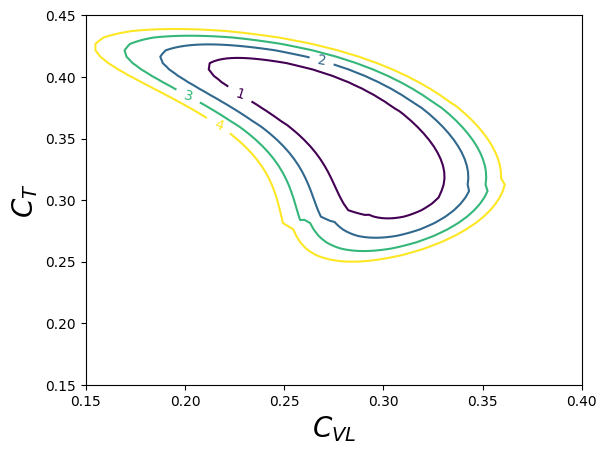}& 
     \includegraphics[width=0.3\textwidth]{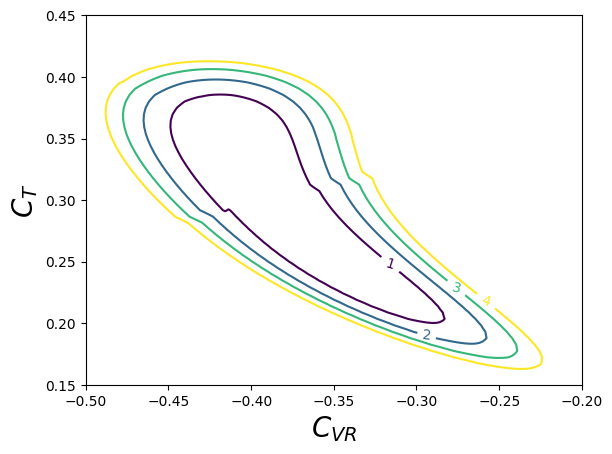}  
    \end{tabular}
    \caption{Fit to the WET parameters to the $R_{D^*}$ and $R_{J/\psi}$ ratios at $1\, \sigma$ (purple), $2\, \sigma$ (blue), $3\, \sigma$ (green) and $4\, \sigma$ (yellow).}
    \label{fig:fit_RDRJ}
\end{figure}
In order to better understand the impact of NP on $R_{J/\psi}$ and its relation to $R_{D^*}$, we perform a statistical fit of the three Wilson coefficients $C_{VR}$, $C_{VL}$ and $C_T$ including both observables. As mentioned before, the contribution of the scalar coefficients are negligible. The fit is performed by numerical minimization of the statistic test $\chi^2$ given by
\begin{equation}
    \chi^2_\mathrm{fit} = \frac{1}{2} \sum_{i, j} [\mathcal{O}_i^\mathrm{exp}-\mathcal{O}_i^\mathrm{th}(\{C\})] \,\mathcal{C}_{ij}^{-1} \,[\mathcal{O}_j^\mathrm{exp}-\mathcal{O}_j^\mathrm{th}(\{C\})]\,,
    \label{eq:chifit}
\end{equation}
where $\mathcal{O}_i^\mathrm{exp}$ and
$\mathcal{O}_i^\mathrm{th}(\{C\})$ are the experimental measurement and
theoretical prediction for the $i$-th observable, respectively, and
$\mathcal{C}_{ij}$ is the correlation matrix. The result of the fit is
represented in Figure~\ref{fig:fit_RDRJ}, by choosing two-dimensional
contours of the Wilson coefficients with the rest of the parameters as
fixed in the best fit point. The scalar contributions $C_{SL}$ and
$C_{SR}$ are suppressed and the fit is constrained by the vector Wilson
coefficients $C_{VL}$ and $C_{VR}$ and the tensor coefficient $C_T$. The
values obtained for each coefficient are: 
\begin{align}
    C_{VL} &= 0.296 \pm 0.167\,,&\quad C_{VR} &= -0.409 \pm 0.089\,,&\quad C_T &= 0.341 \pm 0.386\,.
    \label{eq:match_bctaunu}
\end{align}

The predictions for $R_{D}, R_{D^*}$ and $R_{J/\psi}$ observables and their uncertainties in the best fit point are
\begin{equation}
    R_{D} = 0.342^{+0.039}_{-0.039}\,, \qquad R_{D^*} = 0.288^{+0.022}_{-0.022}\,, \qquad R_{J/\psi} = 0.345^{+0.026}_{-0.029}\,,
\end{equation}
compatible with the experimental results within $1\,\sigma$. 

\subsection{\texorpdfstring{\boldmath{$B\to K^{(*)} \nu \bar{\nu}$}}{B K nu nu} decays}
\label{sec:eft_bknunu}

In the WET, the $B\to K^{(*)} \nu \bar{\nu}$ processes receive contributions from the effective operators appearing in the Lagrangian~\cite{Bause:2023mfe}
\begin{equation}
    \mathcal{L}_{b\to s \nu\bar{\nu}} = -\frac{4G_F}{\sqrt{2}}\frac{\alpha_\mathrm{em}}{4\pi}\sum_{\nu, \nu'}\left[(C_L^{\mathrm{NP}\,\nu\nu'}+C_L^{\mathrm{SM}\,\nu\nu'})\,O_L^{\nu\nu'}+C_R^{\nu\nu'}\,O_R^{\nu\nu'}\right]+\mathrm{h.c.}\,,
\end{equation}
where the semileptonic four-fermion operators are
\begin{equation}
    O_L^{\nu\nu'}=(\bar{s}\gamma_\mu P_L b)(\bar{\nu}\gamma^\mu P_L \nu')\,,\qquad\qquad O_R^{\nu\nu'}=(\bar{s}\gamma_\mu P_R b)(\bar{\nu}\gamma^\mu P_L \nu')\,,
\end{equation}
being $P_{R,L} = (1 \pm \gamma_5)/2$. The SM matching is $C_L^{\mathrm{SM}\,\nu\nu'} = V_{tb}V_{ts}^* \,X_\mathrm{SM} \,\delta_{\nu\nu'}$, where $V_{ij}$ are the elements of the CKM matrix, and $X_\mathrm{SM} = -12.64 \pm 0.15$~\cite{Bause:2023mfe,Brod:2021hsj}. The branching ratios in the WET are calculated as
\begin{eqnarray}
\mathrm{BR}(B^+\to K^+\nu\bar{\nu}) &=& A_+^{BK} \sum_{\nu,\nu'}|(C_L^{\mathrm{NP}\,\nu\nu'}+C_L^{\mathrm{SM}\,\nu\nu'})+C_R^{\nu\nu'}|^2\,, \\
\mathrm{BR}(B^0 \to K^{*0}\nu\bar{\nu}) &=& A_+^{BK^*} \sum_{\nu,\nu'}|(C_L^{\mathrm{NP}\,\nu\nu'}+C_L^{\mathrm{SM}\,\nu\nu'})+C_R^{\nu\nu'}|^2\nonumber\\
&+& A_-^{BK^*} \sum_{\nu,\nu'}|(C_L^{\mathrm{NP}\nu\nu'}+C_L^{\mathrm{SM}\nu\nu'})-C_R^{\nu\nu'}|^2\,,
\end{eqnarray}
where $A_+^{BK} = (5.66\pm 0.17)\times10^{-6}$~\cite{Bause:2023mfe} and $A_-^{BK^*} = (8.88 \pm 1.08)\times 10^{-6}$, $A_+^{BK^*} = (2.00 \pm 0.29) \times 10^{-6}$~\cite{Bause:2021cna}.

The tree-level match between the relevant SMEFT and the WET Wilson coefficients is~\cite{Jenkins:2017jig}
\begin{eqnarray}
    C_L^{\mathrm{NP}\,\nu\nu'} &=& -\frac{\sqrt{2}\pi}{G_F \Lambda^2 \alpha_\mathrm{em}}\left(C_{\ell q(1)}^{\nu\nu' 23}-C_{\ell q(3)}^{\nu\nu' 23}+\delta_{\nu\nu'}C_{\varphi q(1)}^{23}+\delta_{\nu\nu'}C_{\varphi q(3)}^{23}\right)\,,\nonumber\\
    C_R^{\nu\nu'} &=& -\frac{\sqrt{2}\pi}{G_F \Lambda^2 \alpha_\mathrm{em}}\left(C_{\ell d}^{\nu\nu' 23}+\delta_{\nu\nu'}C_{\varphi d}^{23}\right)\,.
    \label{eq:match_bsnunu}
\end{eqnarray}

\begin{figure}
    \centering
    \includegraphics[width=0.4\linewidth]{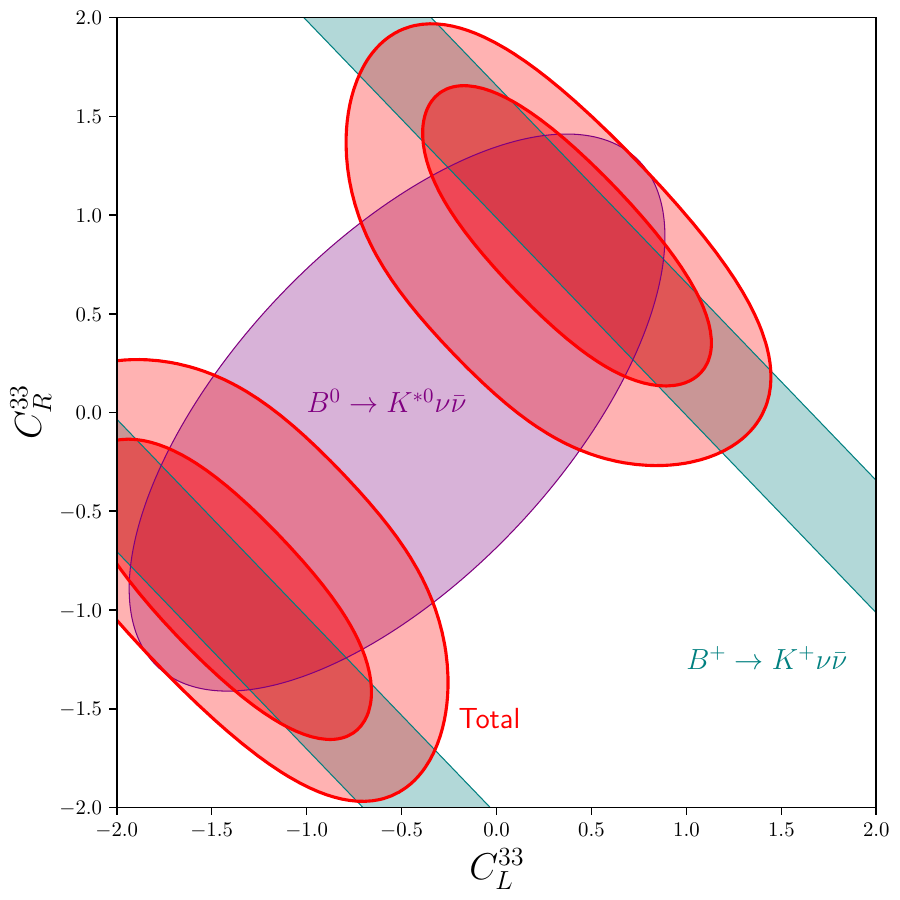}
    \caption{Fit to the WET parameters for $B\to K^{(*)}\nu\bar{\nu}$ branching ratios. The teal areas correspond to $ \mathrm{BR}(B^+\to K^+\nu \bar{\nu}) = (2.3\pm0.7)\times10^{-5}$ and the purple area to $\mathrm{BR}(B^0 \to K^{*0}\nu\bar{\nu}) < 1.8\times 10^{-5}$, while the red areas are the best fit at $1\,\sigma$ and $2\,\sigma$.}
    \label{fig:CL33-CR33}
\end{figure}
When performing a fit that considers only the parameters $C_L^{33}$ and $C_R^{33}$ (third generation of neutrinos), we observe that the resulting values are located at the intersection between two parallel bands, which occur at $C_L^{33}+C_R^{33} = \mathrm{constant}$, consequence of the constraints imposed by the observable $B^+\rightarrow K^+\nu\bar{\nu}$; and the ellipse that corresponds to the constraints imposed by the observable $B^0\rightarrow K^{0*}\nu\bar{\nu}$ as shown in Figure~\ref{fig:CL33-CR33}. The two minima are located at 
\begin{equation}
    (C_L^{33},\, C_R^{33}) = (-1.4\pm0.6,\, -0.9\pm0.6) \qquad \mathrm{and} \qquad (C_L^{33},\, C_R^{33}) = (0.4\pm0.6,\,0.9\pm0.6)\,
\end{equation}
and in both minima the predictions for the observables are
\begin{equation}
    \mathrm{BR}(B^+\to K^+\nu\bar{\nu}) = (2.2\pm0.7)\times 10^{-5},\qquad \mathrm{BR}(B^0\to K^{*0}\nu\bar{\nu}) = (1.3\pm0.2)\times 10^{-5}\,,
\end{equation}
compatible with the experimental measurements. Our results are in agreement with  the results of~\cite{Bause:2023mfe} for their cLFC scenario. If only couplings to left-handed quarks are allowed, the best fits are obtained for $C_L^{33} = 0.9 \pm 0.3$ or $-2.0\pm0.3$, corresponding to $\mathrm{BR}(B^+\to K^+\nu\bar{\nu}) = (1.5\pm0.6)\times 10^{-5}$ and $\mathrm{BR}(B^0\to K^{*0}\nu\bar{\nu}) = (1.8\pm1.0)\times 10^{-5}$. Therefore, we can conclude that $C_L^{33}$ alone can reduce the tension to the $1\,\sigma$ level, while the addition of $C_R^{33}$ is necessary to fully describe the experimental data.

\section{Global fits}
\label{sec:GF}

In this section we will update and extend the results obtained
in~\cite{Alda:2021rgt}, which used a subset of the SMEFT operators first
proposed by~\cite{Feruglio:2017rjo,Cornella:2018tfd,Feruglio:2018jnu}. The
dimension-6 Lagrangian in the ``Warsaw-down'' basis takes the
form~\cite{Alda:2021rgt} 
\begin{equation}
    \mathcal{L}_\mathrm{NP} = \frac{\lambda^\ell_{ij} \lambda^q_{kl}}{\Lambda^2} [C_1 (\bar{\ell}_i \gamma_\mu \ell_j)(\bar{q}_k \gamma^\mu q_\ell) + C_3 (\bar{\ell}_i \gamma_\mu\tau^I \ell_j)(\bar{q}_k \gamma^\mu \tau^I q_\ell)]\,,
    \label{eq:LagNP}
\end{equation}
where $\ell_i$ and $q_i$ are the $SU(2)_L$ doublets for the $i$-th generation of leptons and quarks, respectively, $\Lambda$ is the energy scale where the UV theory is matched to the SMEFT, $C_1$ and $C_3$ are the Wilson coefficients for singlet and triplet interactions, with the identification $C_{\ell q(1)}^{ijkl} = C_1 \lambda^\ell_{ij} \lambda^q_{kl}$ and $C_{\ell q(3)}^{ijkl} = C_3 \lambda^\ell_{ij} \lambda^q_{kl}$, being $\lambda^\ell$ and $\lambda^q$ the matrices that determine the flavour structure of the NP interactions. We will fix the value $\Lambda = 1\,\mathrm{TeV}$, which assumes that the NP effects can be observed in flavour physics experiments. Flavour-changing charged currents can only be mediated by triplet interactions, and therefore $C_3$ is necessary to describe the $R_{D^{(*)}}$ and $R_{J/\psi}$ anomalies. Flavour-changing neutral currents, on the other hand, are mediated by both $C_1$ and $C_3$, in particular the contributions to $b\to s \ell^+ \ell^-$ are proportional to $C_1+C_3$, while $b\to s \nu\bar{\nu}$ receives tree-level contributions from $C_1-C_3$. Finally, the matrices $\lambda$ are a reflection of the flavour structure of the UV theory. One simple case, proposed by~\cite{Feruglio:2017rjo}, considered that the UV theory only affected one generation of fermions in a certain interaction basis, which needs to be rotated to the mass basis. In this case, the matrices must be hermitian, idempotent ($\lambda^2 = \lambda$) and $\mathrm{tr}\,\lambda = 1$ and, as explained in~\cite{Alda:2021rgt}, can be parameterized by two complex numbers, $\alpha^{\ell,q}$ and $\beta^{\ell,q}$, in the following way:
\begin{equation}
    \lambda^{\ell,q} = \frac{1}{1+|\alpha^{\ell,q}|^2+|\beta^{\ell,q}|^2} \begin{pmatrix}
        |\alpha^{\ell,q}|^2 & \alpha^{\ell,q}\bar{\beta}^{\ell,q} & \alpha^{\ell,q} \\
        \bar{\alpha}^{\ell,q}\beta^{\ell,q} & |\beta^{\ell,q}|^2 & \beta^{\ell,q} \\
        \bar{\alpha}^{\ell,q} & \bar{\beta}^{\ell,q} & 1
    \end{pmatrix}\,. \label{eq:lambdamatrix}
\end{equation}

In order to be consistent in our analysis, we must not restrict ourselves to the ratios $R_{D^{(*)}}$ and $R_{J/\psi}$. We have performed the fits including all observables implemented by \texttt{smelli} version 2.3.3~\cite{Aebischer:2018iyb} (nearly 600 in total, covering Higgs decays, diboson production, semileptonic charm decays, lepton flavour violation, electroweak precision tests, and various $B$-physics channels) plus $R_{J/\psi}$\footnote{We implemented the observable $R_{J/\psi}$ using the form factors in~\cite{Harrison:2020gvo}.}. 

This broad inclusion is methodologically essential for several reasons. First, the Langrangian~\eqref{eq:LagNP} directly contributes to many other physical observables, for example the previously mentioned $\mathrm{BR}(B\to K^{(*)}\nu\bar{\nu})$ and even Lepton Flavour violating processes, like $B\to K e^\pm \mu^\mp$, due to the off-diagonal terms of $\lambda^\ell$. This way, while the dominant sensitivity to our parameters $(C_1, C_3, \beta^q)$ comes from $B$-physics observables, the full set constrains the flavor mixing encoded in $\lambda^q$ after rotating to the mass basis, breaks approximate degeneracies in the parameter space, and validates the stability of the preferred region. Second, the running of the Renormalization Group equations causes the mixing of the operators in Lagrangian~\eqref{eq:LagNP} with other SMEFT operators. For example, the insertion of $O_{\ell q(1)}$ ($O_{\ell q (3)}$) is necessary for the one loop corrections to $O_{\varphi q(1)}$ and $O_{\varphi \ell(1)}$ ($O_{\varphi q(3)}$ and $O_{\varphi \ell(3)}$) operators that are relevant for interactions between $Z$ bosons and fermions. Third, including observables from multiple sectors provides crucial null tests: improvements in $b\to c\tau\nu$ or $b\to s\nu\bar{\nu}$ channels should not reappear as new tensions in light-lepton channels, charm physics, or electroweak precision measurements.

\begin{table}
    \centering
    \begin{tabular}{|c|c|c|c|}\hline
         & Scenario I & Scenario II  & Scenario III\\\hline\hline
       $C_1$ & $-0.11^{+0.03}_{-0.04}$  & $-0.12\pm 0.03$ & $-0.205\pm 0.015$ \\\hline
       $C_3$ & $-0.11^{+0.03}_{-0.04}$  & $-0.12\pm0.03$ & $-0.12^{+0.02}_{-0.01}$ \\\hline
       $\alpha^\ell$ & -- & $0.0 \pm 0.07$ & -- \\\hline
       $\beta^\ell$ & $0.00\pm 0.02$ & $0.000 \pm 0.014$ & -- \\\hline
       $\alpha^q$ & -- & $-0.076$ & -- \\\hline
       $\beta^q$ & $0.78^{+1.22}_{-0.36}$ & $0.85^{+1.05}_{-0.6}$ & $0.64^{+1.36}_{-0.24}$ \\\hline\hline
       Pull & $5.71\,\sigma$ & $5.51\,\sigma$ & $6.25\,\sigma$ \\\hline
       $\Delta \chi^2_\mathrm{SM}$ & 39.8 & 43.12 & 46.66 \\\hline
       $p$-value & $1.2\times 10^{-8}$ & $3.5\times 10^{-8}$ & $4.1\times 10^{-10}$ \\\hline
    \end{tabular}
    \caption{Results of the best fits to the rotation parameters and the coefficients $C$ in Scenario I, II and III.}
    \label{tab:fit_results}
\end{table}
 The goodness of each fit is evaluated with its difference of $\chi^2$ with respect to the SM, $\Delta\chi^2_\mathrm{SM}=\chi^2_\mathrm{SM}-\chi^2_\mathrm{fit}$, being $\chi^2_\mathrm{fit}$ as defined in~(\ref{eq:chifit}). Besides, the package \texttt{smelli} computes the differences of the logarithms of the likelihood function as $\Delta \log L = -\frac{1}{2} \Delta \chi^2_\mathrm{SM}$ and, in order to compare two fits $A$ and $B$, we use the pull between them in units of $\sigma$, defined as~\cite{Capdevila:2018jhy} 
\begin{equation}
\mathrm{Pull}_{A \to B} = \sqrt{2} \mathrm{Erf}^{-1}[F(\Delta \chi^2_A - \Delta \chi^2_B; n_B - n_A )]\,,
\end{equation}
where $\mathrm{Erf}^{-1}$ is the inverse of the error function, $F$ is the cumulative distribution function of the $\chi^2$ distribution and $n$ is the number of degrees of freedom of each fit. 

For comparison with our previous results, the SM input parameters are chosen to be the same as in~\cite{Alda:2021rgt} and we have considered the two scenarios defined in this paper. In addition, motivated by the tension in the new $B^+\to K^+\nu\bar{\nu}$ measurement at Belle II~\cite{Belle-II:2023esi} together with the alleviation of the tension in the $R_{K^{(*)}}$ ratios, we have introduced a new scenario called Scenario III:
\begin{itemize}
\item \textbf{Scenario I:} Includes only mixing between the second and third generations through $\beta^\ell$ and $\beta^q$, that is, $\alpha^\ell = \alpha^q = 0$ and $C_1 = C_3\equiv C$.
\item \textbf{Scenario II:} The only assumption is $C_1 = C_3 \equiv C$. Mixing with both first and second generation leptons through $\alpha^\ell$, $\beta^\ell$, $\alpha^q$ and $\beta^q$ is considered.
\item \textbf{Scenario III:} $C_1$ and $C_3$ are independent parameters, and we do not consider mixing in the lepton sector, that is $\alpha^\ell = \beta^\ell = 0$. 
\end{itemize}

Scenarios I and II assumed that $C_1=C_3\equiv C$ because at that time, the $R_{K^{(*)}}$ observables presented deviations as compared to their SM predictions, while the constraints set by the $B\to K^{(*)}\nu\bar{\nu}$ decays were quite stringent. However, at present the experimental situation has reversed, with the tension in the $R_{K^{(*)}}$ ratios disappearing~\cite{LHCb:2022qnv} and a new anomaly appearing in $B^+ \to K^+ \nu \bar{\nu}$, pointing towards different NP contributions to $C_1$ and $C_3$. Scenario III assumes that $C_1$ and $C_3$ are independent parameters. Additionally, since there are no longer discrepancies in muonic observables, we do not consider mixing in the lepton sector ($\alpha^\ell = \beta^\ell = 0$); we are also discarding mixing to the first quark generation since it played no role in our previous fits. In all three scenarios, only real values for the fit parameters are considered. 

The results of the best fits to the rotation parameters and the coefficients $C$ are summarized in Table~\ref{tab:fit_results}. Scenarios I and II present a similar best fit point, with $C \sim -0.12$, $\beta^q=0.8$ and the rest of the rotation parameters compatible with zero. The $\Delta \chi^2_\mathrm{SM}$ is also similar, with $\Delta \chi^2_\mathrm{SM}=39.8$ in Scenario I and $\Delta \chi^2_\mathrm{SM}=43.1$ in Scenario II, which results in Scenario I having a better pull because there are less degrees of freedom, i.e. it is a simpler hypothesis. As anticipated, Scenario III results in a much better fit as compared to Scenarios I and II, since it requires the minimum number of parameters needed to describe simultaneously the anomalies in $R_{D^{(*)}}$ and $B\to K \nu \bar{\nu}$, with $\Delta \chi^2_\mathrm{SM} = 46.66$ and a pull of $6.25\,\sigma$. 

In Scenario III, since $\alpha^\ell = \beta^\ell = 0$, there is no mixing in the leptonic sector, and the NP only affects the third generation, i.e., $\lambda^\ell_{33} = 1$ and $\lambda^\ell_{ij} = 0$ for all the other entries. Meanwhile, with a fit value of $\beta^q$ close to one, there is a large mixing between the second and third quark generations, resulting in $\lambda^q_{22} = 0.29$, $\lambda^q_{33} = 0.71$ and $\lambda^q_{23} = \lambda^q_{32} = 0.45$. This mixing structure thus provides large contributions to $b\to s \,\nu_\tau \bar{\nu}_\tau$ and $b\to c \,\tau^- \bar{\nu}_\tau$ transitions while leaving $b\to s \,e^+ e^-$ and $b\to s \,\mu^+ \mu^-$ unaltered at tree level. 

\begin{figure}
    \centering
    \begin{tabular}{ccc}
         \includegraphics[width=0.3\textwidth]{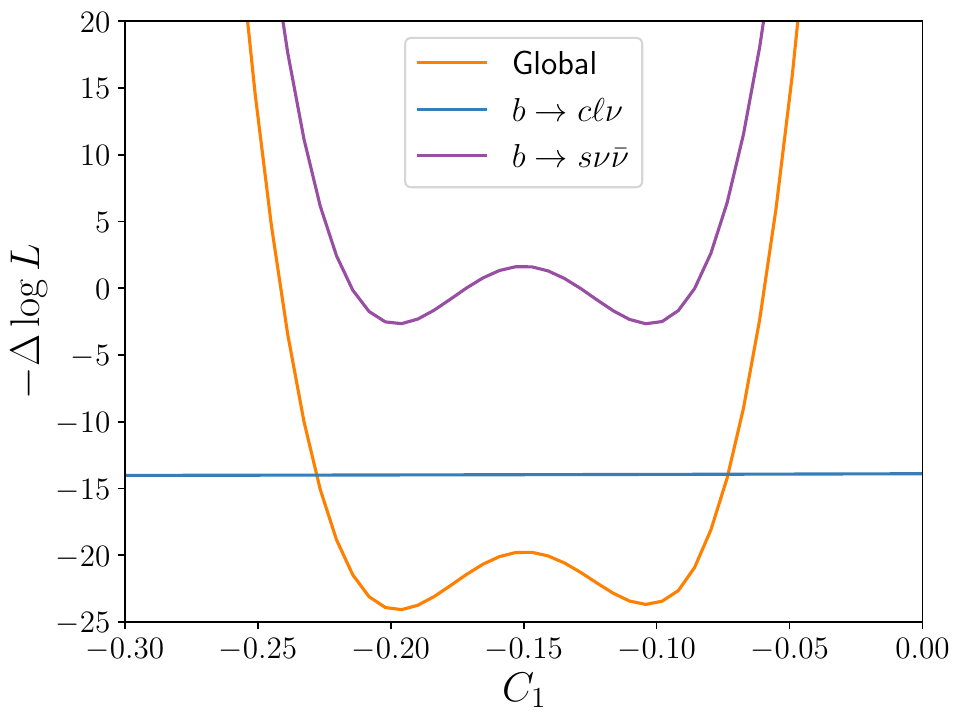} & \includegraphics[width=0.3\textwidth]{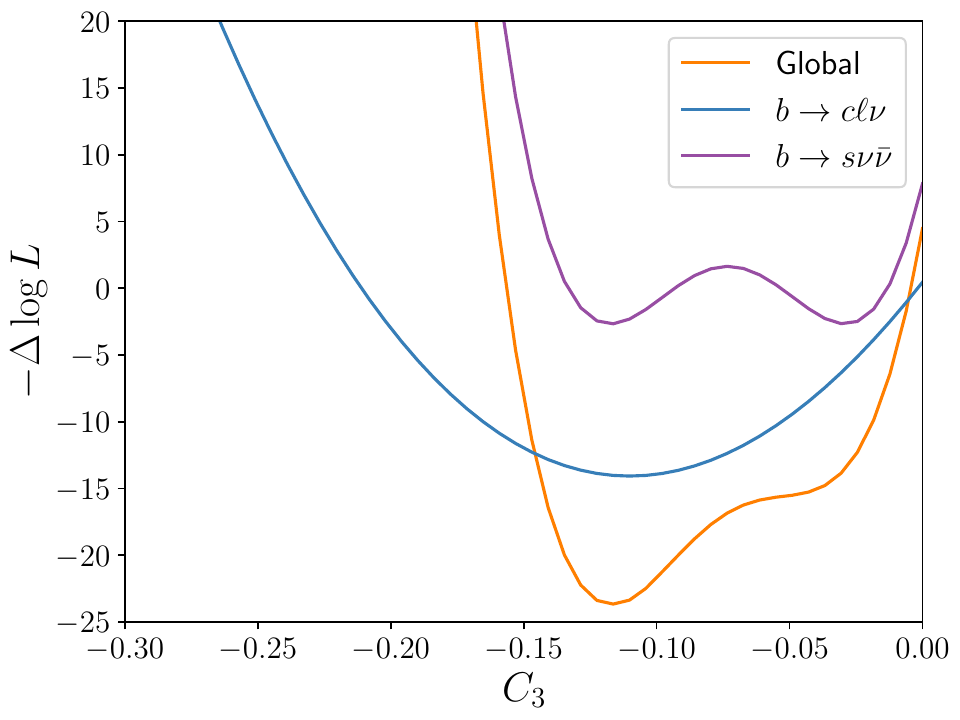} & \includegraphics[width=0.3\textwidth]{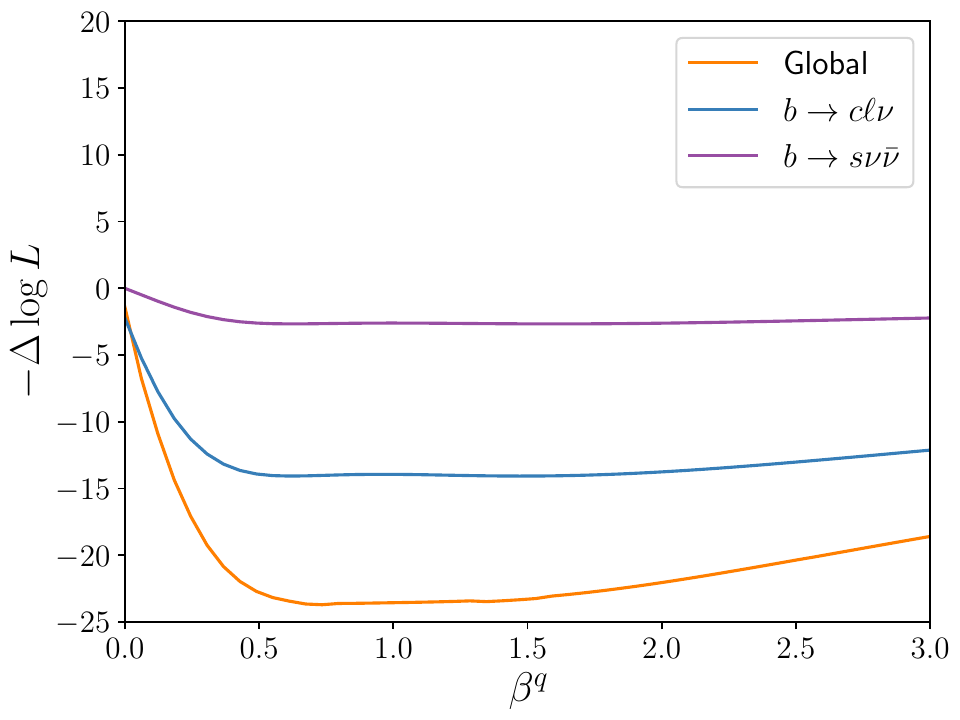}\\
         (a) & (b) & (c) 
    \end{tabular}
    \caption{Negative of the logarithm of the likelihood function when varying (a) $C_1$ (b) $C_3$ and (c) $\beta^q$ with respect to their values at the best fit for Scenario III.}
    \label{fig:scIII_likelihood1D}
\end{figure}
\begin{figure}
    \centering
    \begin{tabular}{ccc}
         \includegraphics[width=0.3\textwidth]{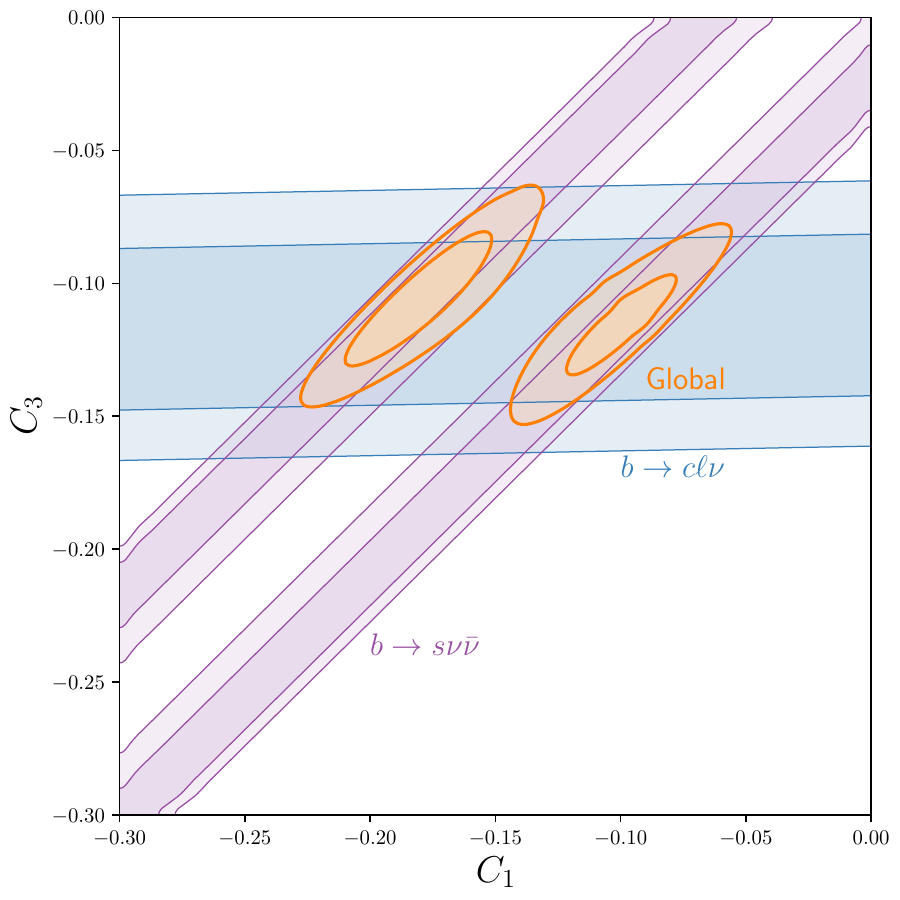} & \includegraphics[width=0.3
         \textwidth]{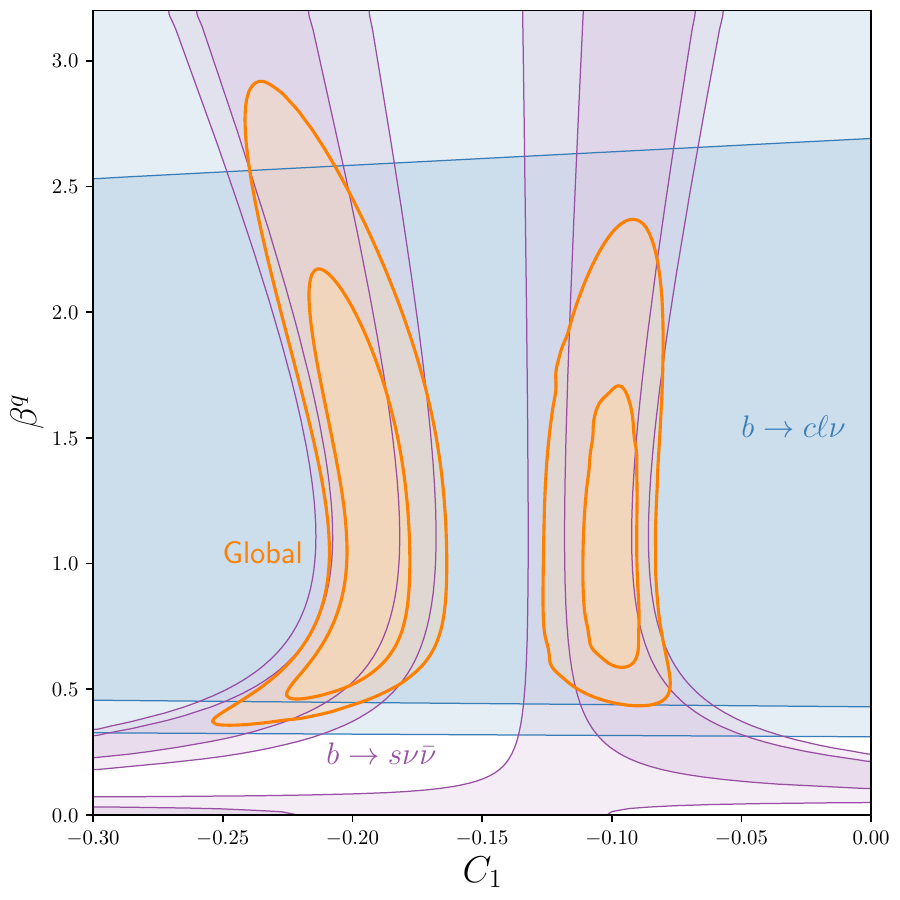} & \includegraphics[width=0.3\textwidth]{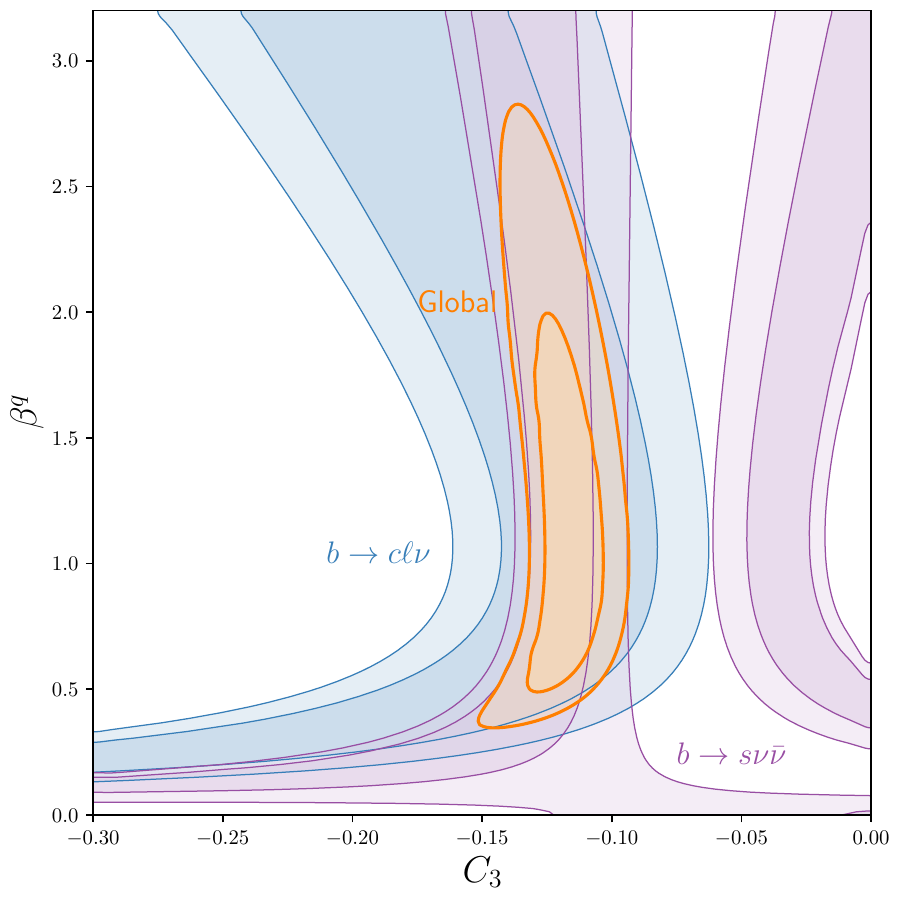}\\
         (a) & (b) & (c) 
    \end{tabular}
    \caption{Regions of constant likelihood in Scenario III for (a) $C_1$ and $C_3$ plane (b) $C_1$ and $\beta^q$ plane and (c) $C_3$ and $\beta^q$ plane, with the rest of parameters as in the best fit point of Scenario III.}
    \label{fig:scIII_likelihood}
\end{figure}
The role of each parameter of the fit in constraining each class of observables is shown in Figures~\ref{fig:scIII_likelihood1D} and~\ref{fig:scIII_likelihood}. The one-dimensional slices of the likelihood function for Scenario III is included in Figure~\ref{fig:scIII_likelihood1D}. The global fit (in orange) consists of 593 observables. However, it can be seen that it is largely dominated by the two classes of observables that we have been discussing in the previous sections: $b\to c\ell \nu$ (in blue) and $b\to s \nu\bar{\nu}$ (in purple). In particular, the $b\to s\nu\bar{\nu}$ observables determine the likelihood function in the direction of the parameter $C_1$, both $b\to c\ell\nu$ and $b\to s\nu\bar{\nu}$ contribute to the likelihood function in the direction of $C_3$, and $b\to c\ell\nu$ observables present the most important contribution to the likelihood in the direction of $\beta^q$. A similar picture emerges in Figure~\ref{fig:scIII_likelihood} when the contours of constant likelihood are considered. As indicated by the matching conditions in~\eqref{eq:matching} and \eqref{eq:match_bsnunu}, the Wilson coefficient $C_{VL}$ relevant for the $b\to c\ell\nu$ decays depends on $C_{\ell q(3)}^{3323} = C_3 \lambda^q_{23}$ while the coefficient $C_L^{33}$ of the $b\to s \nu\bar{\nu}$ observables depends on both $C_{\ell q(1)}^{3323} = C_1 \lambda^q_{23}$ and $C_{\ell q(3)}^{3323} = C_3 \,\lambda^q_{23}$.

After applying the running of the Renormalization Group equation and the matching conditions (see~\cite{Alda:2021rgt} for details), the Wilson coefficients relevant for the $b\to c \,\tau^- \bar{\nu}_\tau$ and $b\to s \,\nu_\tau \bar{\nu}_\tau$ processes at the scale $\mu = m_b$ are
\begin{equation}
    C_{VL} = 0.089\,, \qquad\qquad C_L^{33} = 0.61\,,
\end{equation}
while the other coefficients in Eq.~\eqref{eq:matching} and Eq.~\eqref{eq:match_bsnunu} do not receive contributions from Scenario III. There is also a contribution to the Wilson coefficient $C_9^\ell$ which enters the $b\to s \,\ell^+ \ell^-$ decays for $\ell = e,\,\mu$~\cite{Alda:2021rgt,Crivellin:2018yvo}. Since this contribution is generated radiatively from $C_9^\tau$, it is lepton flavour universal for the light flavours, $C_9^e = C_9^\mu = -0.58$, and consequently it does not affect the $R_{K^{(*)}}$ ratios, which remain at their SM values  $R_{K^{(*)}} \approx 1$. However, this contribution does enter in the branching ratios of $B\to K^{(*)}\mu^+\mu^-$ and $B_s\to \phi \,\mu^+ \mu^-$.

\begin{figure}
    \centering
    \includegraphics[width=0.6\textwidth]{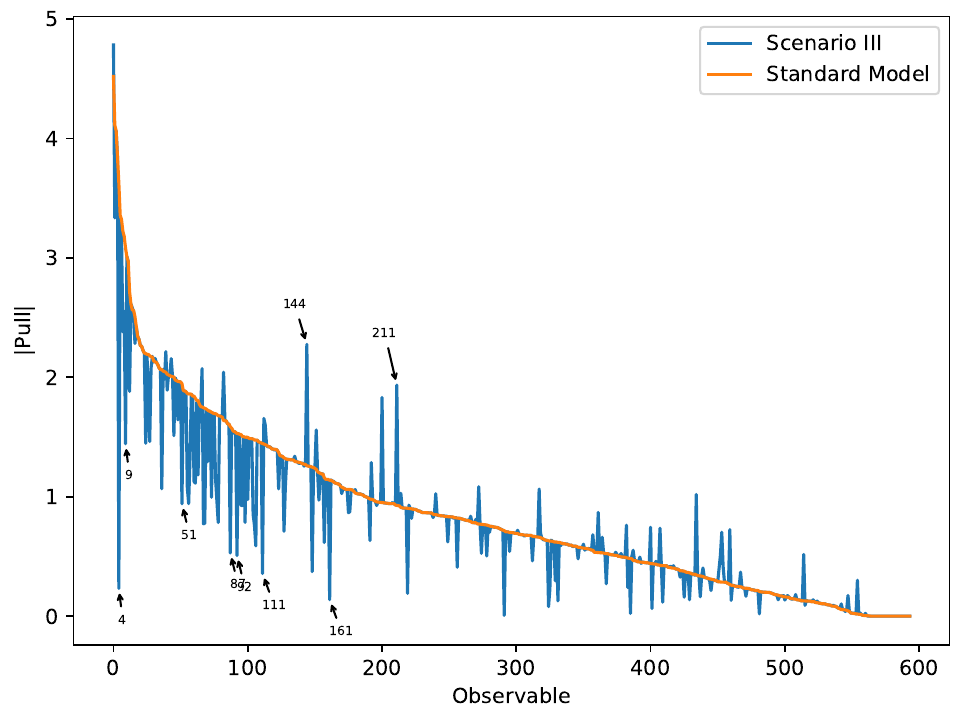}
    \caption{Pull of the observables considered for the global fit in the SM (orange line) and Scenario III (blue line). The observables are ordered by decreasing SM pull, and the ones in which the SM and Scenario III differ more than $1\,\sigma$ are specially marked.}
    \label{fig:pulls}
\end{figure}
We have obtained the predictions for all the observables included in the fit, as well as the pull in the SM and in Scenario III (NP pull). Figure~\ref{fig:pulls} shows the comparison of the pull of each observable with respect to their experimental value in the SM (orange line) and in the Scenario III (blue line). Here the observables in which the SM and Scenario III differ more than $1\,\sigma$ are highlighted. We observe that the highlighted observables are all related to the $B$-meson anomalies: $R_{D^*}^\ell$ (observable 4) and $R_D^\ell$ (observable 92) improve by $3.3\,\sigma$ and $1\,\sigma$ respectively; $\mathrm{BR}(B^+ \to K^+ \nu\bar{\nu})$ (observable 9) improves by $1.7\,\sigma$ and $\mathrm{BR}(B^+ \to K^{*+} \nu\bar{\nu})$ (observable 161) by $1\,\sigma$ while $\mathrm{BR}(B^0 \to K^{*0} \nu\bar{\nu})$ (observable 144) gets worse by $1\,\sigma$. And finally, regarding the branching ratios of $b\to s\mu^+\mu^-$ at high $q^2$, $\mathrm{BR}(B^\pm\to K^\pm \mu^+ \mu^-)$ in the $[16-17]\,\mathrm{GeV}^2$ (observable 51) and  $[15-22]\,\mathrm{GeV}^2$ (observable 87) bins, as well as $\mathrm{BR}(B_s^\pm\to \phi^\pm \mu^+ \mu^-)$ in the $[15-19]\,\mathrm{GeV}^2$ bin (observable 111),  improve by around $1\,\sigma$ while  $\mathrm{BR}(B^\pm\to K^\pm \mu^+ \mu^-)$ in the $[17-18]\,\mathrm{GeV}^2$ bin (observable 211) gets worse also by $1\,\sigma$. As pointed out in Section~\ref{sec:eft_bknunu}, it is not possible to describe $B^+ \to K^+ \nu\bar{\nu}$ and $B^0 \to K^{*0} \nu\bar{\nu}$ decays only with left-handed effective operators, consequently resulting in a worse pull for $B^0 \to K^{*0} \nu\bar{\nu}$. As for the $B^+\to K^+ \mu^+\mu^-$ differential branching ratios, both LHCb~\cite{LHCb:2014cxe} and CMS~\cite{CMS:2024syx} find some tension between the  $[17-18]\,\mathrm{GeV}^2$ bin, which is measured above the SM prediction, and the rest of the bins in the $q^2 \geq 15\,\mathrm{GeV}^2$, which are below their SM prediction, and our fit reproduces said tension. 

\subsection{Comparison with previous results}

The inclusion of the new experimental measurements at LHCb for $R_{K^{(*)}}$~\cite{LHCb:2022qnv,LHCb:2022vje} has significantly impacted our global fit, as compared to previous results presented in~\cite{Alda:2021rgt}. In the previous fit, we found that a sizable contribution to $\alpha^\ell$ was needed to describe the $R_{K^{(*)}}$ anomaly via NP affecting the electron decay mode. Now that the experimental measurements for $R_{K^{(*)}}$ are compatible with the SM predictions, the current fit no longer requires large deviations in $\alpha^\ell$. In consequence, there is no mixing in the leptonic sector, with $\lambda^\ell_{33} \approx 1$ affecting the tau decays (required to describe LFUV in $b\to c\tau\nu_\tau$ processes), while $\lambda^\ell_{ij} \approx 0$ for $ij \neq 33$. In the quark sector, the central values found by the fit present no significant changes as compared to the previous one, with large NP effects in $\beta^q$ needed to describe the anomalies in flavour-changing $b\to c$ charged currents. However, we observe that $\beta^q$ presents a larger allowed range, including even regions with $\beta^q > 1$, that is, with larger NP effects in the second generation than in the third generation. In the previous fit, $\beta^q$ was constrained by the LFV decays $B\to K^{(*)} \mu e$, which received NP contributions through $C \lambda^q_{23} \lambda^\ell_{12} \sim C \beta^q \alpha^\ell \beta^\ell$. With the new fit preferring values of $\alpha^\ell$ much closer to zero, the impact of $\beta^q$ in the LFV observables is decreased, and instead, $\beta^q$ is controlled mainly by the $b \to c \ell \nu$ decays. 

\begin{figure}
    \centering
    \begin{tabular}{cc}
         \includegraphics[trim= 0 30 0 0 ,clip,width=0.4\textwidth]{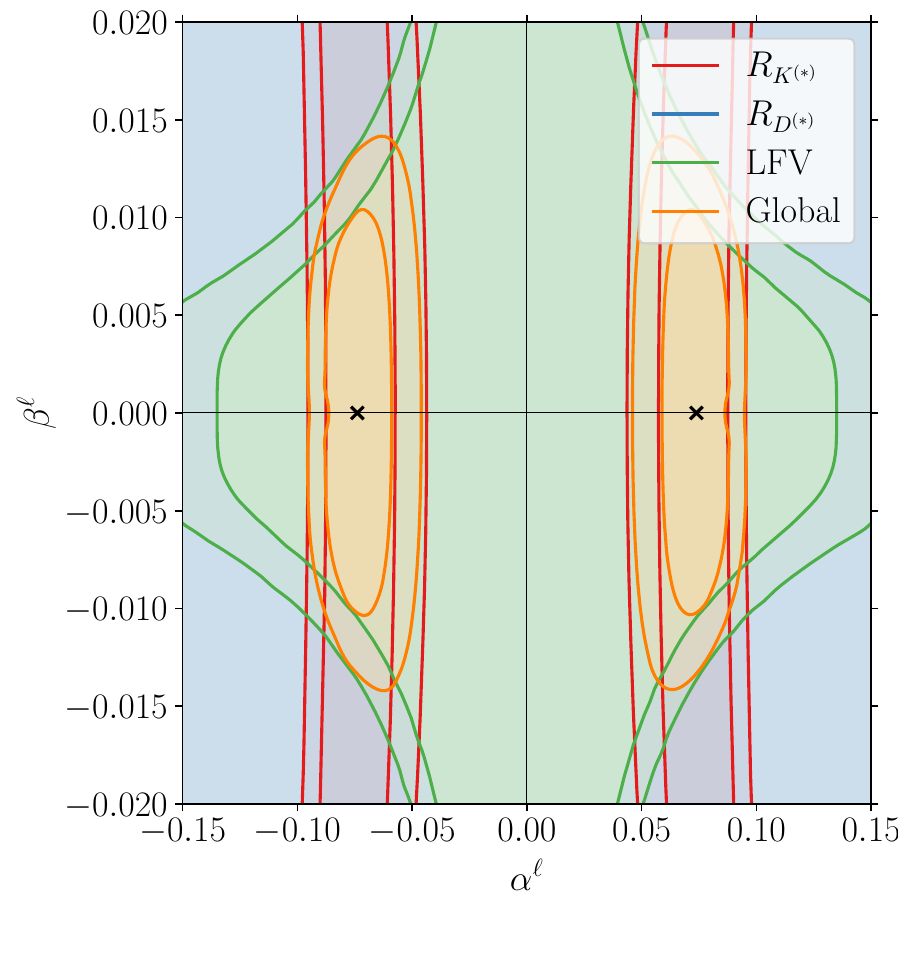} & \includegraphics[width=0.4\textwidth]{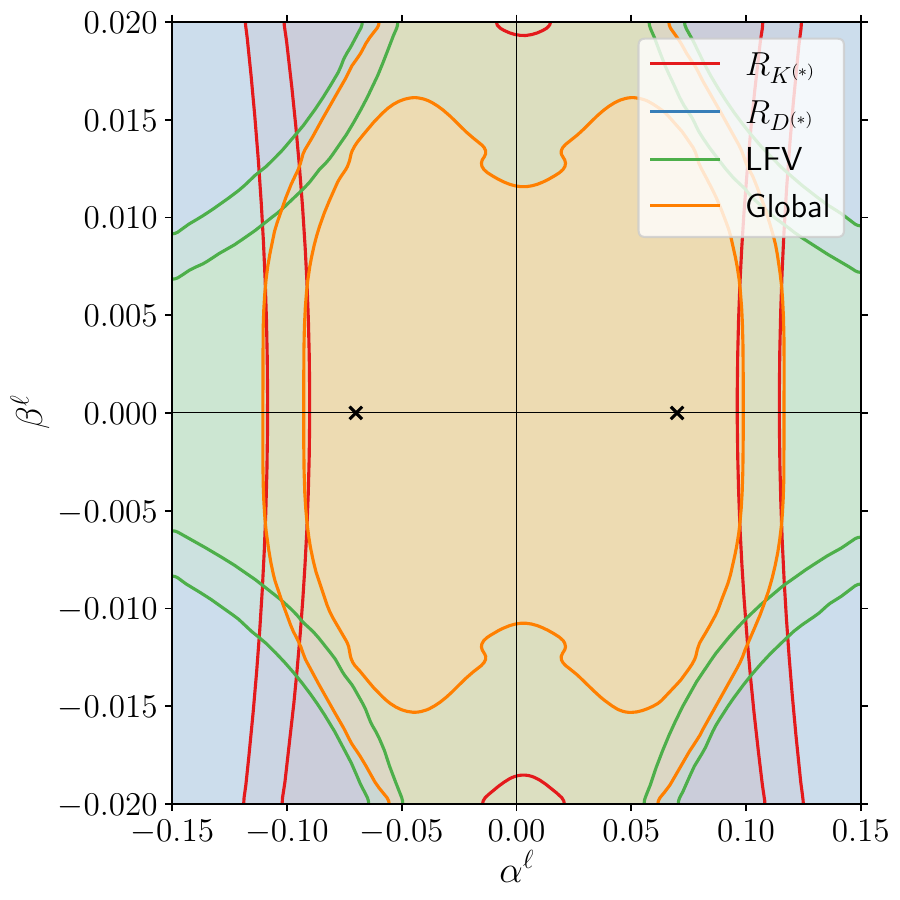} \\
         (a) & (b) 
    \end{tabular}
    \caption{Regions of constant likelihood for $\alpha^\ell$ and $\beta^\ell$ parameters in Scenario II as (a) in the previous work~\cite{Alda:2021rgt} and (b) in the current work. }
    \label{fig:likelihood_plots}
\end{figure}
This is illustrated in Figure \ref{fig:likelihood_plots}, where the two-dimensional section of the likelihood function for the parameters $\alpha^\ell$-$\beta^\ell$ in Scenario II, with the rest of the parameters set at the best fit point of this scenario, is compared between the previous fit (left) and the new fit (right). In~\cite{Alda:2021rgt} we found that the better fit, with the inclusion of previous discrepancies on $R_K$ and $R_{K^*}$ results, corresponded to Scenario II. However, the present experimental measurements of these ratios do not have discrepancies with the SM predictions anymore. In fact, it can be seen that the main difference is the region allowed for the $R_{K^{(*)}}$ observables, inside the red lines: in the previous fit, the allowed region corresponded to $0.04 \leq |\alpha^\ell| \leq 0.10$, which was not compatible with the SM value $\alpha^\ell = 0 $ owing to the experimental anomalies at that moment. On the other hand, the allowed region in the current fit is $0 \leq |\alpha^\ell| \leq 0.12$, which includes the SM point since the anomalies are gone. The global fit, in orange, follows the same logic.

\begin{figure}
    \centering
    \begin{tabular}{cc}
         \includegraphics[height=0.33\textwidth]{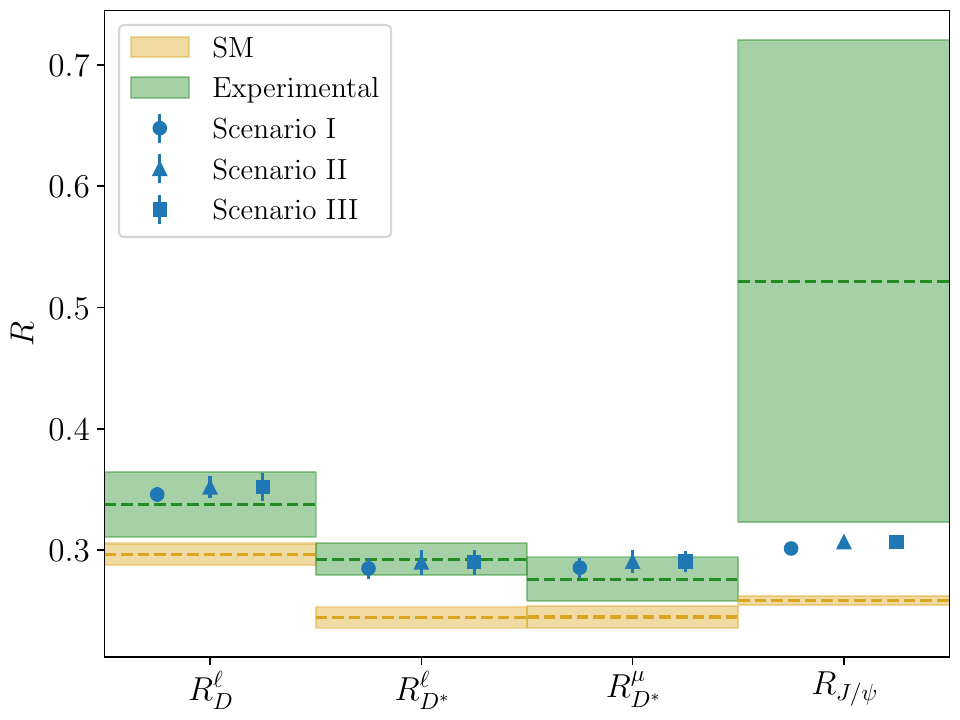} & \includegraphics[height=0.33\textwidth]{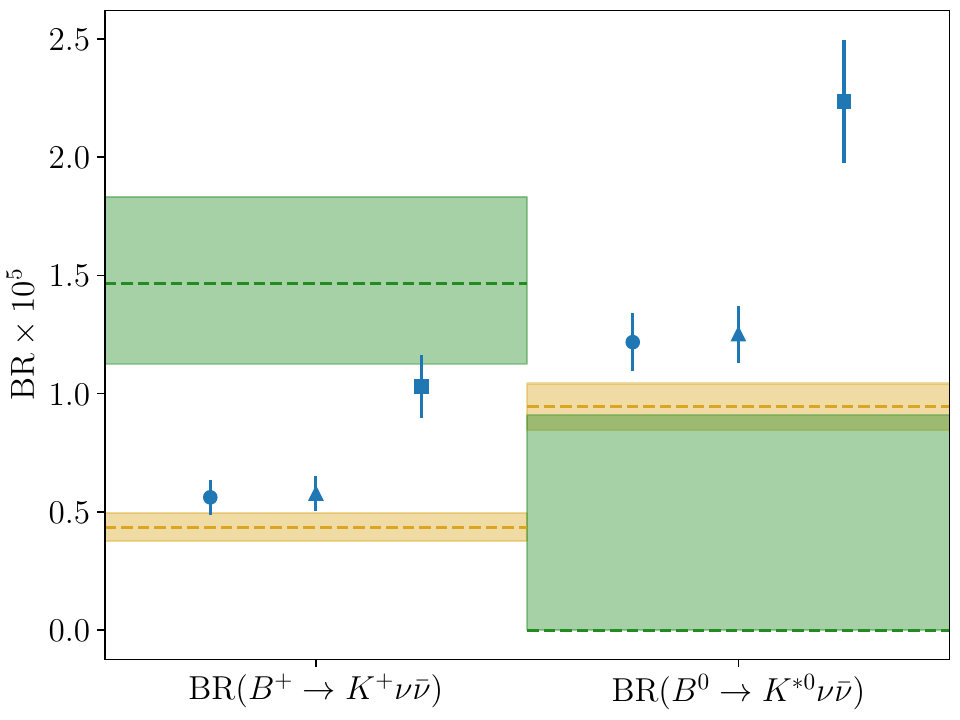}  \\
         (a) & (b) 
    \end{tabular}
    \caption{Predictions for selected (a) $b\to c \ell \nu$ LFU ratios and (b) $b\to s \nu\bar{\nu}$ branching ratios. The predictions in Scenarios I, II and III and their expected error are represented with blue markers. The SM prediction and its $1\,\sigma$ error are represented by the yellow regions. The experimental measurement and its $1\,\sigma$ error are represented by the green regions.}
    \label{fig:RKRD}
\end{figure}
The predictions for $R_D$, $R_{D^*}$ and $R_{J/\psi}$ observables in the best fit points for Scenarios I, II and III are shown in Figure~\ref{fig:RKRD}(a). $R_D$ and $R_{D^*}$ exhibit very similar values in both scenarios II and III (as $\alpha^{l,q}$ does not affect these predictions), and these fitted values align well with experimental results. The current fit is largely dominated in the $b\to c\ell\nu$ sector by the $R_{D^{(*)}}$ ratios, as their uncertainty is much smaller than in the case of $R_{J/\psi}$. It would be therefore desirable to improve the measurements of $R_{J/\psi}$ in order to obtain an independent handle of the $b\to c\ell \nu$ sector and further clarify the present anomaly. 

Figure~\ref{fig:RKRD}(b) includes the predictions for $B^+\to K^+\nu\bar{\nu}$ and $B^0\to K^{*0}\nu\bar{\nu}$ branching ratios, not included in our previous analysis. In both cases we can see how the scenario I and II, in which the coefficients $C_1=C_3$ were not independent, fail to correctly capture the NP contribution, yielding values very close to the SM predictions. In contrast, Scenario III manages to improve the results in the charged sector, although the results in the neutral sector are worsened.

For comparison, the values obtained for $b\to s\,\ell^+\ell^-$, $b\to c\,\ell \,\nu$ and $b\to s \,\nu\,\bar{\nu}$ observables in Scenario II are presented in Table~\ref{tab:comp_RKRD}. For each observable, the significance of the deviations with respect to the experimental values, in units of $\sigma$, is provided. The ones in the column ``Previous fit'' are referred to the experimental measurements that were available in Ref.~\cite{Alda:2021rgt}, while the ones in the column ``Current fit'' correspond to the updated values derived in this work. In the case of the $R_{K^{(*)}}$ ratios, the current fit improves the pull for $R_{K^{*0}}^{[1.1, 6.0]}$ and $R_{K^{*0}}^{[0.045, 1.1]}$, the later by almost $2\,\sigma$. At the same time, the pull for $R_{K^+}^{[1.1,\ 6.0]}$ has increased to $1.1\,\sigma$. The reason for this change is that this is the only bin where some of the tension with the SM prediction still survives, and our fit reproduces this tension. The updated measurements of $R_{D^{(*)}}$ and the inclusion of the observable $R_{J/\psi}$, with its large uncertainty, have a lesser impact on our fit. However, for the $R_{D^{(*)}}$ ratio that includes the decay mode to electrons, $R_{D^*}^\ell$, the current fit represents an improvement over the previous one due to the reduced NP effects in $\alpha^\ell$. 
\begin{table}[]
    \centering
    \begin{tabular}{|c|c|c|}\hline
         Observable & Previous fit & Current fit  \\ \hline
         $R_{K}^{[1.1,\ 6.0]}$ & 0.83438 ($0.28\,\sigma$) &  1.0009 ($1.1\,\sigma$) \\\hline
         $R_{K^{*0}}^{[0.045, 1.1]}$ & 0.88373 ($2.1\,\sigma$) & 0.92444 ($0.026\,\sigma$) \\\hline
         $R_{K^{*0}}^{[1.1, 6.0]}$ & 0.84052 ($1.4\,\sigma$) & 0.99609 ($0.41\,\sigma$) \\\hline\hline
         $R_D^\ell$ & 0.35166 ($0.17\,\sigma$) & 0.35131 ($0.51\,\sigma$) \\\hline
         $R_{D^*}^\ell$ & 0.2898 ($0.41\,\sigma)$ & 0.28951($0.23\,\sigma)$ \\\hline
         $R_{D^*}^\mu$ & 0.29041 ($0.75\,\sigma)$ & 0.28997 ($0.77\,\sigma)$ \\\hline\hline
         $R_{J/\psi}$ & - & 0.30608 ($1.1\,\sigma)$ \\\hline\hline
         $\mathrm{BR}(B^+ \to K^{+}\nu\bar{\nu})$ & $5.7715\times 10^{-6}$ ($1.1\,\sigma$) & $9.7026\times 10^{-6}$ ($1.4\,\sigma$)\\\hline
         $\mathrm{BR}(B^0 \to K^{*0}\nu\bar{\nu})$ & $1.2523\times10^{-5}$ ($1.5\,\sigma$)& $2.1038\times10^{-5}$ ($2.3\,\sigma$) \\\hline
    \end{tabular}
    \caption{Comparison of the predictions for $b\to s\ell^+\ell^-$, $b\to c\ell \nu$ and $b\to s \nu\bar{\nu}$ observables between our previous fit and the current fit in Scenario III. The significance of the predictions in the column ``Previous fit'' are referred to the experimental measurements that were available in Ref.~\cite{Alda:2021rgt}, while the significances in the column ``Current fit'' correspond to the present values.}
    \label{tab:comp_RKRD}
\end{table}

The impact of the new measurement and the reduced NP effects in $\alpha^\ell$ can also be noticed in other $b\to s \,e^+ e^-$ observables, for example in the inclusive branching ratio $\mathrm{BR}(B\to X_s \,e^+ e^-)$, that has changed from $1.3\,\sigma$ to $1.6\,\sigma$ in the $[14.2,\,25.0]\,\mathrm{GeV}^2$ bin and from $0.23\,\sigma$ to $0.74\,\sigma$ in the $[1.0,\,6.0]\,\mathrm{GeV}^2$ bin. Also in LFU ratios, like other bins of $R_{K^{(*)}}$ that were measured by Belle~\cite{BELLE:2019xld}, for example $R_{K^0}^{[4.0,\,8.12]}$ changing from $0.86\,\sigma$ to $1.3\,\sigma$ and $R_{K^+}^{[4.0,\,8.12]}$ from $1\,\sigma$ to $0.59\,\sigma$. By contrast, $b\to s\,\mu^+\mu^-$ observables like the angular observables ${P'}_4$ and ${P'}_5$ or the branching ratios $\mathrm{BR}(B\to K^{*}\mu^+\mu^-)$ and $\mathrm{BR}(B_s\to \phi\,\mu^+\mu^-)$ remain unchanged between both fits, because the value for $\beta^\ell$ has not changed.

\subsection{Machine Learning analysis}
\label{sec:ML}

In this section we detail the Machine Learning (ML) methodology employed in this work, building on the framework established in~\cite{Alda:2021rgt}. While our scenarios contain between 3 and 6 fit parameters (a modest dimensionality for which standard sampling techniques are in principle adequate), we employ an ML-based approach for several compelling reasons that prove essential for this analysis. The confidence intervals extracted from our global fit exhibit pronounced non-Gaussian features: the likelihood surfaces develop curved ridges in the $(C_1, C_3, \beta^q)$ parameter space, weakly constrained directions emerge from approximate degeneracies, and the mixing of parameters across different scales ($C_1 \sim \mathcal{O}(0.1)$ vs. $\beta^q \sim \mathcal{O}(1)$) complicates straightforward grid-based sampling. 
These features arise naturally from the interplay between multiple observables with different sensitivities to the Wilson coefficients, and accurately capturing them is crucial for reliable uncertainty propagation to physical predictions.

Given the relatively modest dimensionality of the parameter space, one might expect standard sampling techniques like Markov Chain Monte Carlo (MCMC) to be sufficient. However, the ML emulator offers distinct advantages for this specific analysis due to the likelihood's complex topology. While MCMC is efficient for posterior integration, standard algorithms can struggle to mix efficiently in parameter spaces featuring the narrow, curved ridges present in our global fit. The tree-based regressor adaptively partitions these non-Gaussian boundaries. Furthermore, once trained, the emulator allows for the instantaneous generation of millions of sample points. This computational speed is crucial for producing high-resolution confidence contours and performing the SHAP sensitivity analysis, tasks that would be computationally prohibitive if dependent on running fresh, high-statistics MCMC chains for each projection.

ML, specifically gradient-boosted regression trees implemented via XGBoost (eXtreme Gradient Boosting) algorithm~\cite{Chen:2016btl}, provides an efficient and robust solution tailored to these challenges. The choice of tree-based methods over alternative ML architectures such as neural networks is deliberate and advantageous for our specific use case. Regression trees partition the parameter space in an adaptive manner by sequentially applying binary splits. This approach directs computational resources towards regions of high curvature or rapid variation without assuming global smoothness or continuity. This property is particularly valuable in our context, where the likelihood function exhibits narrow ridges, sharp transitions near best-fit points, and plateau-like regions corresponding to weakly constrained parameter combinations. 

Neural networks, by contrast, are universal function approximators that learn smooth interpolations through gradient descent on continuous activation functions. While powerful for high-dimensional problems with abundant training data, they typically require significantly larger training sets and careful regularization to avoid over-smoothing the fine structure present in the likelihood landscape. Moreover, neural networks are prone to over-fitting in low-dimensional regimes without extensive hyperparameter tuning, and their training can be sensitive to initialization and learning rate schedules.

Boosted decision trees, on the other hand, achieve the required accuracy with modest training requirements and provide stable performance under reseeding without extensive tuning. Using the XGBoost algorithm we train an ensemble of regression trees capable of approximating the log-likelihood function of our fit. We build a training sample of 10,500 parameter points with their corresponding likelihood values and split it into two parts: $84\%$ of the points for training and $16\%$ of the points for validation of the model. We train the ML model with a learning rate of $0.03$ and allow early stopping after $5$ rounds of stagnant learning. The training process converges after $776$ boosting rounds, indicating that the model has learned the essential structure of the likelihood landscape without over-fitting.

\begin{figure}
    \centering
    \begin{tabular}{cc}
         \includegraphics[width=0.35\textwidth]{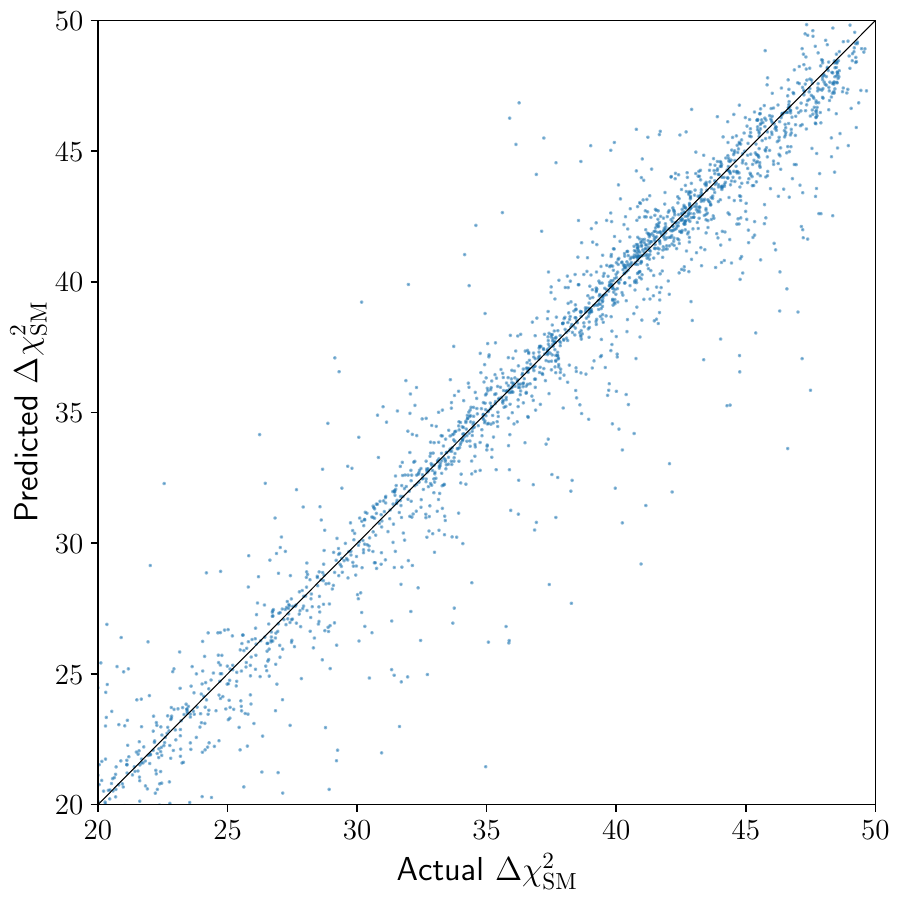} & \includegraphics[width=0.48\textwidth]{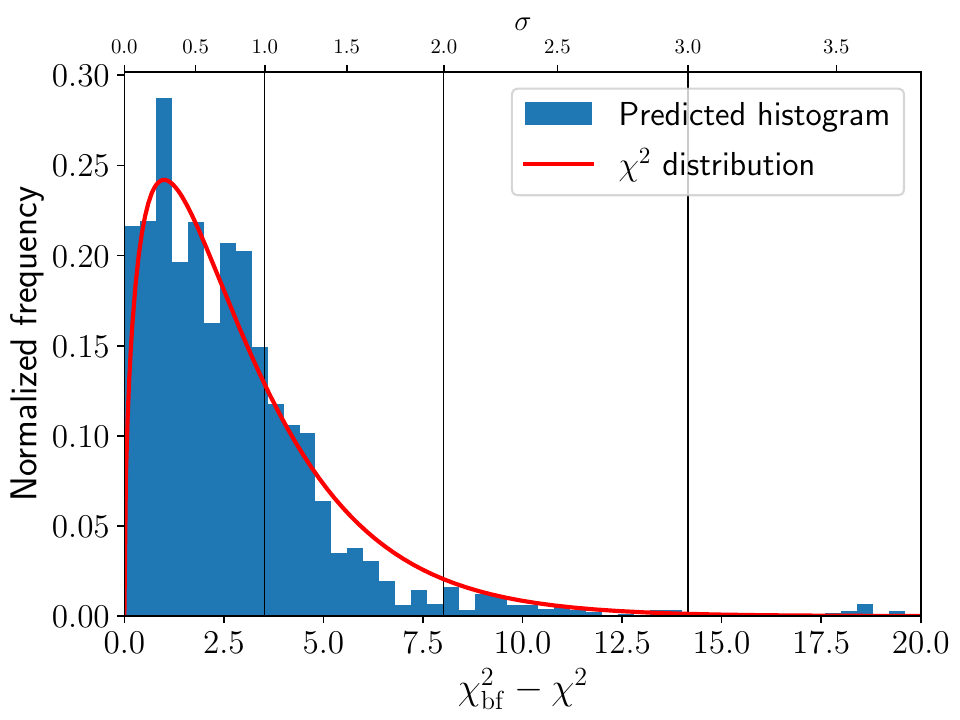} \\
         (a) & (b) 
    \end{tabular}
    \caption{(a) Linear regression between the actual values of the $\Delta\chi^2_\mathrm{bf}$ (horizontal axis) and the values predicted by the XGBoost model (vertical axis) in the validation set in Scenario III. (b) Histogram of the sample of points generated by the ML-based Monte Carlo algorithm in Scenario III, in blue, compared to the $\chi^2$ probability distribution function, in red.}
    \label{fig:reg_scIII}
\end{figure}

The correlation between the actual values of the $\chi^2_\mathrm{SM}$ in the validation dataset and the corresponding predictions by the ML model is included in Figure~\ref{fig:reg_scIII}(a) for Scenario III. The Pearson coefficient for linear correlation is $r=0.96$, indicating an excellent agreement. The histogram for the predicted $\chi^2$ values in Figure~\ref{fig:reg_scIII}(b) shows how, once again, the ML-generated points closely follow the general shape of the $\chi^2$ distribution. 

Furthermore, tree-based models offer superior interpretability through SHAP (Shapley Additive exPlanations) values, which decompose the model output into additive contributions from each input parameter. SHAP analysis reveals not only the global importance of each parameter but also how individual data points drive the fit quality in different regions of parameter space. This diagnostic is particularly valuable for validating that the dominant constraints indeed arise from the expected $B$-physics channels, as we demonstrate below. In summary, while neural networks could in principle approximate our likelihood function, the tree-based approach provides the optimal balance of accuracy, efficiency, robustness, and interpretability for the specific characteristics of our $3-6$ dimensional, non-Gaussian fitting problem with moderate training data availability.

\begin{figure}
    \centering
    \includegraphics[width=0.8\textwidth]{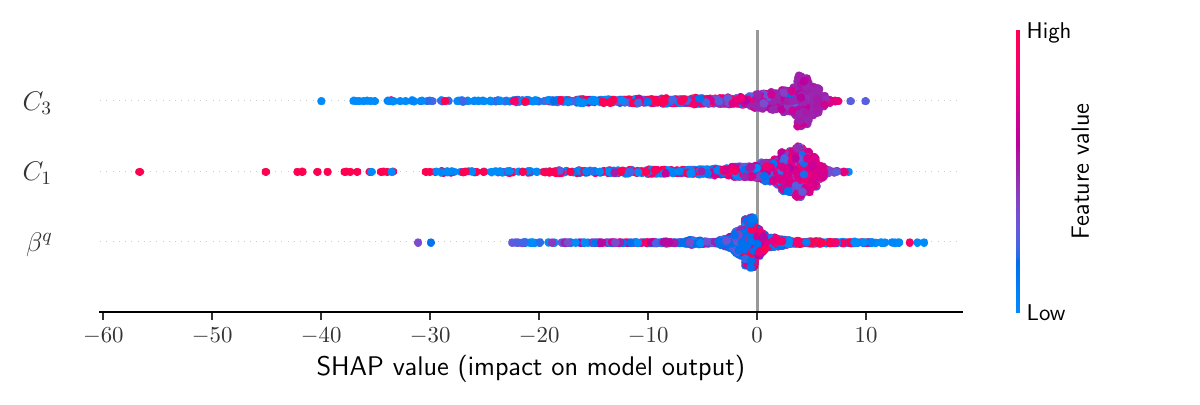}
    \caption{SHAP summary plot for Scenario III showing the impact of each fit parameter ($C_3$, $C_1$, and $\beta^q$) on the model output (likelihood). Each point represents a sample from the parameter space, with the horizontal position indicating the SHAP value (contribution to the log-likelihood) and the color indicating the feature value (red for high, blue for low).}
    \label{fig:SHAP_summary}
\end{figure}
We perform a SHAP analysis to quantify the relative contribution of each parameter to the model likelihood. Figure~\ref{fig:SHAP_summary} summarizes the global feature importance and directional effects within Scenario III. Features are ordered from top to bottom according to their global importance (mean absolute SHAP value). This hierarchy aligns with the underlying physics: the dominant influence of $C_3$ reflects the strong sensitivity of the global fit to charged-current observables such as $R_{D^{(*)}}$, while the second largest contribution from $C_1$ highlights the complementary role of neutral-current processes like $B \to K^{(*)}\nu\bar{\nu}$. The smaller but non-negligible contribution of $\beta^q$ indicates that quark-mixing effects modulate both sectors without dominating the overall likelihood. This consistency between feature importance and the expected physical dominance of charged- and neutral-current operators confirms that the ML approach captures the correct EFT dynamics rather than overfitting numerical noise. The horizontal position of each point represents the SHAP value, corresponding to the impact of that parameter on the log-likelihood, while the color encodes the relative magnitude of the parameter. The plot reveals both the strength and direction of each parameter’s influence on the model outcome: broad distributions correspond to features with large overall impact, whereas dense clusters near SHAP $\sim 0$ mark regions where the model is relatively insensitive to small variations.

\begin{figure}
    \centering
    \begin{tabular}{ccc}  \includegraphics[width=0.3\textwidth]{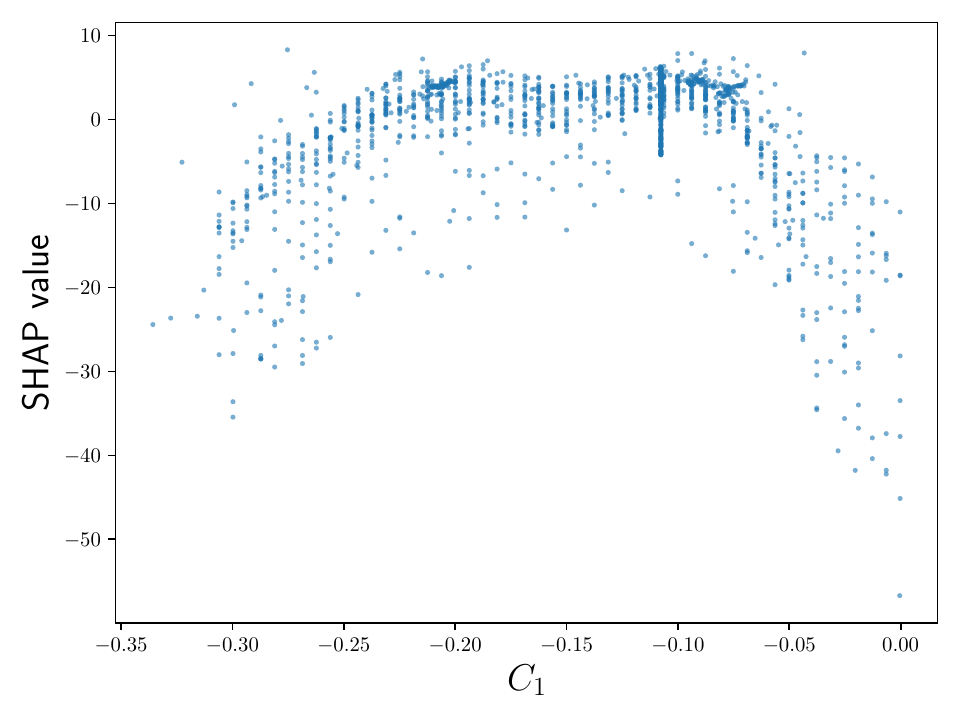} & \includegraphics[width=0.3\textwidth]{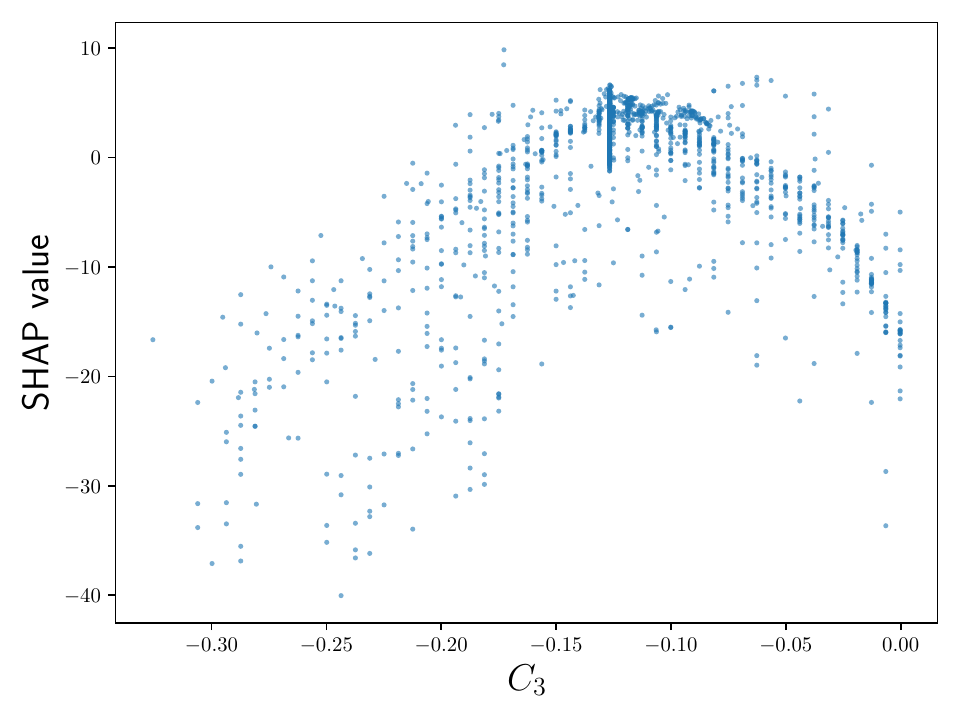} & \includegraphics[width=0.3\textwidth]{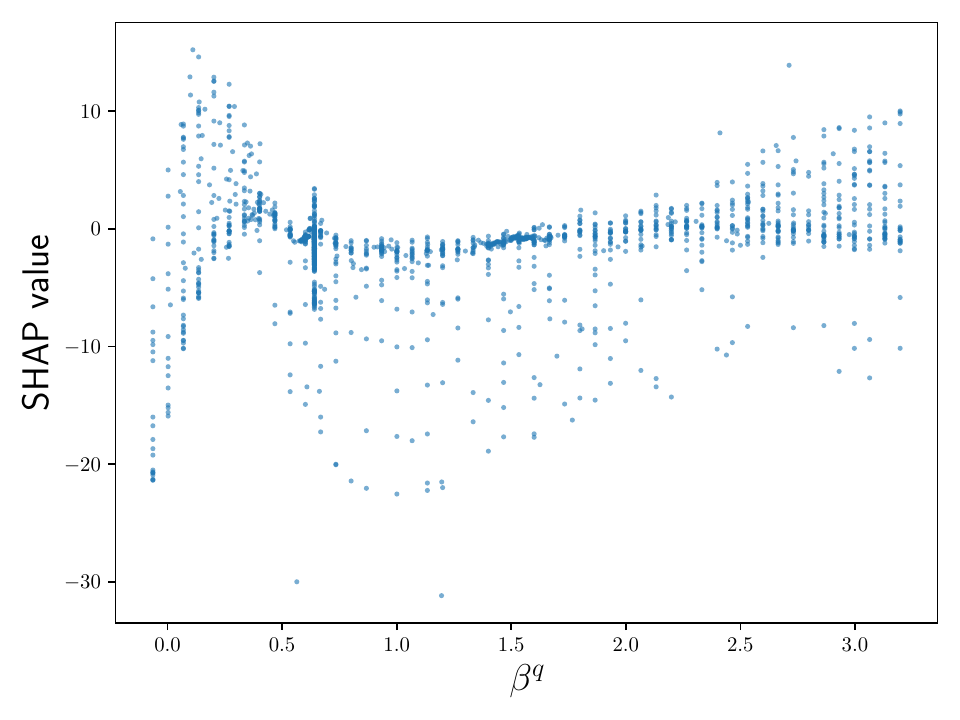}  \\
        (a) & (b) & (c) 
    \end{tabular} 
    \caption{Distribution of SHAP values for each parameter of the fit at the sample of $10500$ generated points: (a) $C_1$, (b) $C_3$, (c) $\beta^q$.}
    \label{fig:SHAP_logL_scIII}
\end{figure}
We can study the behavior of each feature in detail by plotting the logarithm of the likelihood while varying the other parameters, as shown in Figure~\ref{fig:SHAP_logL_scIII}. Note that the parameters follow the shapes obtained in the global fit as presented  in Figure~\ref{fig:scIII_likelihood1D} (with negative sign). The columns of dots visible in the three plots result from an over-representation of the best-fit points in the data sets for $C_1=-0.11$, $C_3=-0.12$ or $\beta^q=0.6$. This effect arises from using a combination of two different data sets: one with randomly generated points across the parameter space, and another with points in a grid used for the elaboration of Figure~\ref{fig:scIII_likelihood}. In the latter, each pair of parameters is sampled while keeping the remaining parameter fixed at its best-fit value. We can see this way how increasing or decreasing affects the SHAP values reproducing the behaviour shown in Figure~\ref{fig:SHAP_summary}.

\begin{figure}
    \centering
    \begin{tabular}{ccc}
         \includegraphics[width=0.3\textwidth]{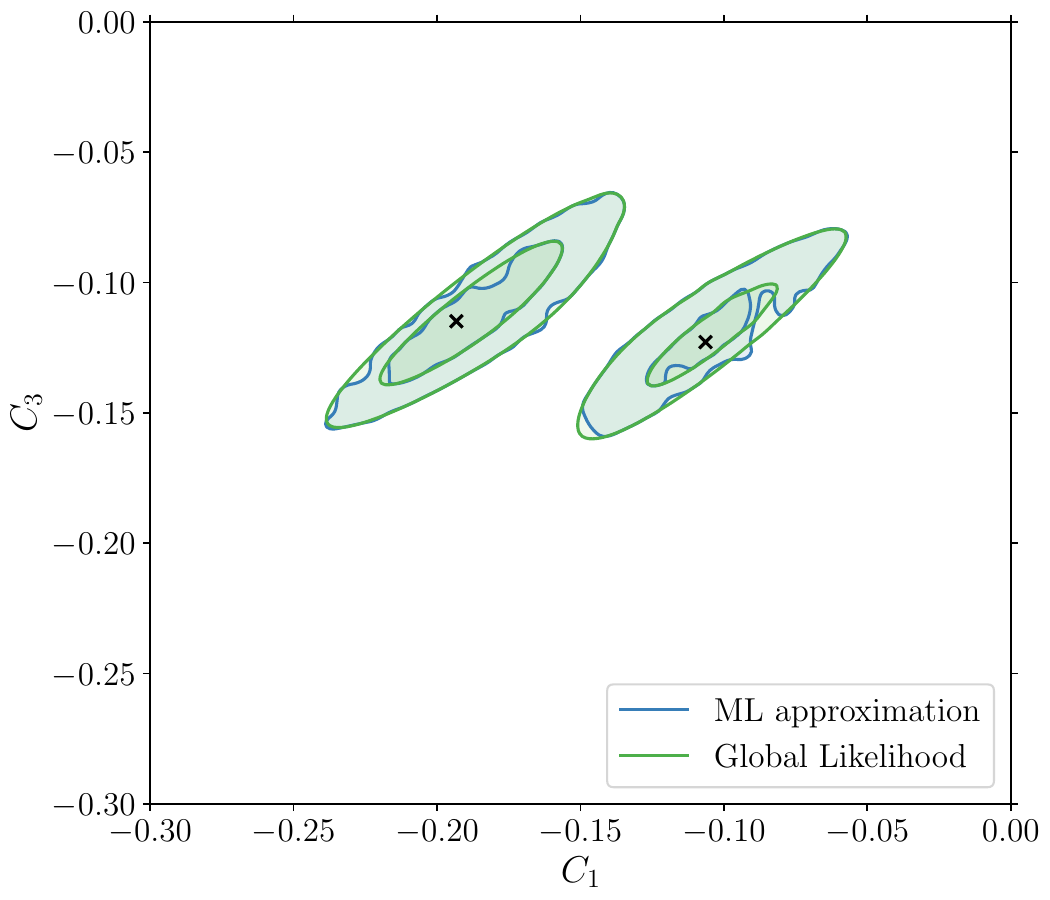} & \includegraphics[width=0.3\textwidth]{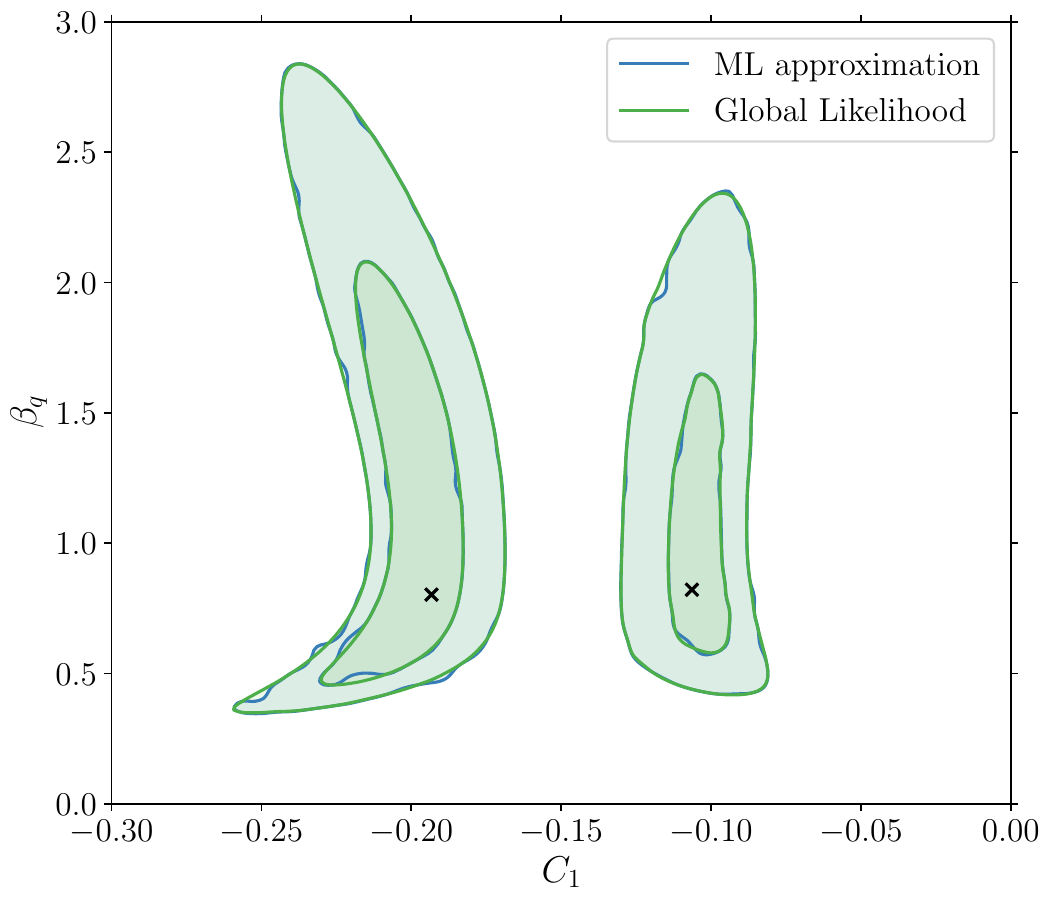} & \includegraphics[width=0.3\textwidth]{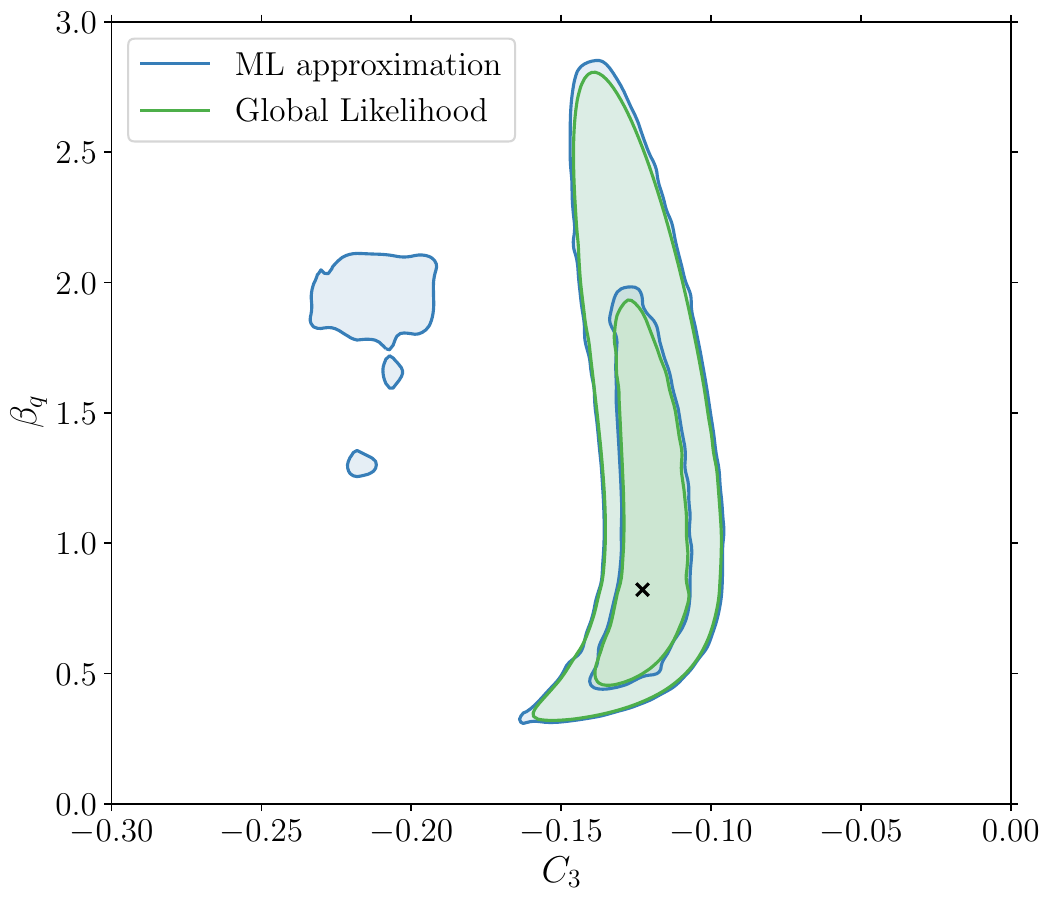} \\
         (a) & (b) & (c) 
    \end{tabular}
    \caption{Comparison between the $1\,\sigma$ and $2\,\sigma$ likelihood contours obtained using the ML approximation (blue) and the actual contours (green) in (a) $C_1$ and $C_3$ plane, (b) $C_1$ and $\beta^q$ plane and (c) $C_3$ and $\beta^q$ plane.}
    \label{fig:ml_contours}
\end{figure}
 We also compare the two-dimensional confidence regions extracted from the ML emulator with those obtained from exact likelihood evaluations. Figure~\ref{fig:ml_contours} shows the $1\,\sigma$ and $2\,\sigma$ contours in the $(C_1, C_3)$, $(C_1, \beta^q)$, and $(C_3, \beta^q)$ planes, with ML-derived contours in blue and exact contours in green. The excellent agreement across all three projections demonstrates that the emulator accurately captures not only the location of the best-fit point but also the shape, orientation, and extent of the confidence regions. This high-fidelity reconstruction validates the ML approach and enables us to use the rapidly-generated samples for detailed correlation studies and uncertainty propagation that would be computationally prohibitive with exact evaluations alone.

 \subsubsection{Correlations between $R_{D^*}$ and $\mathrm{BR}(B^+\to K^+ \nu\bar{\nu})$}

\begin{figure}
    \centering
    \includegraphics[width=0.45\linewidth]{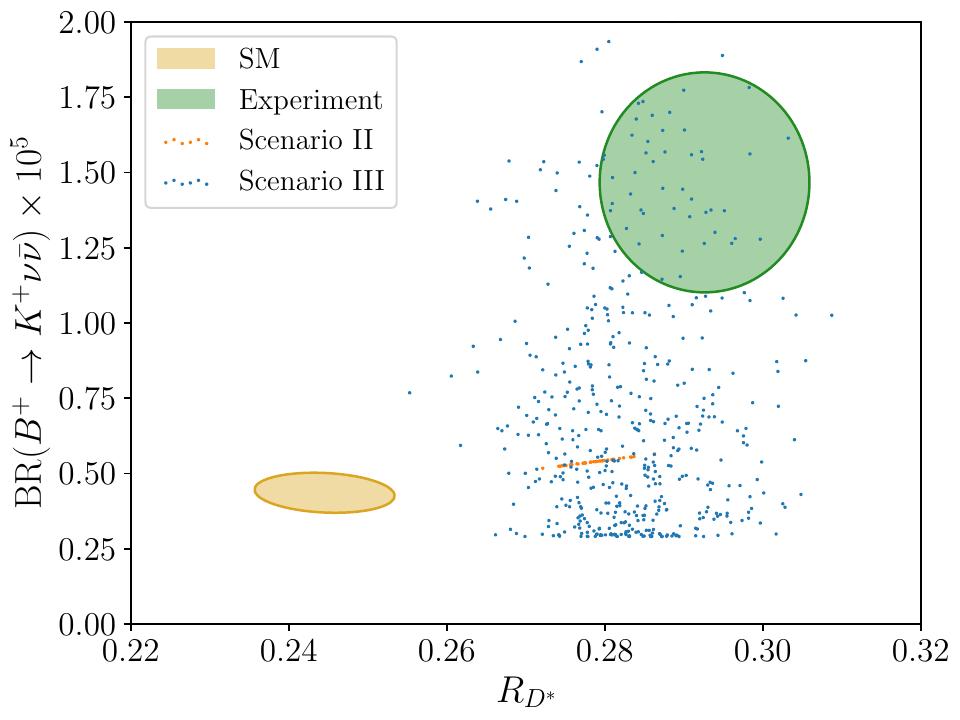}
    \caption{Values of the observables $R_{D^*}$ and $\mathrm{BR}(B^+\to K^+ \nu\bar{\nu})$ in the ML-generated samples for scenario II (orange points) and scenario III (blue points), compared to the SM prediction (yellow region) and experimental measurement (green region).}
    \label{fig:correlation_RDs_BKnunu}
\end{figure}

The inclusion of the new experimental results on $R_{D^*}$ ratios at Belle II~\cite{Belle-II:2024ami}, together with the introduction of our new Scenario III, in which the Wilson coefficients $C_1$ and $C_3$ are treated as independent parameters and lepton-sector mixing is neglected, leads to substantial modifications of results obtained in~\cite{Alda:2021rgt}. In that study, we found an almost-perfect correlation between these two observables. Figure~\ref{fig:correlation_RDs_BKnunu} presents the ML-generated predictions for $R_{D^*}$ and $\mathrm{BR}(B^+\to K^+ \nu\bar{\nu})$ in Scenario II (as considered in the previous work) and Scenario III. As can be seen, the strong correlation observed in Scenario II disappears once Scenario III is considered.

The lack of a consistent correlation between the enhanced $B^+ \to K^+ \nu \bar{\nu}$ branching ratios and $R_{D^*}$ raises important questions about lepton flavor universality and the potential presence of NP phenomena. Further experimental results and theoretical works are necessary to clarify these discrepancies and to understand the underlying physics.

Constructing a UV complete model that reproduces the interplay between $R_{D^*}$ and $\mathrm{BR}(B^+\to K^+\nu\bar{\nu})$ is out of the scope of this work. Nevertheless, several frameworks proposed in the literature account for the $R_{D^{(*)}}$ anomalies while enhancing $\mathrm{BR}(B^+\to K^+\nu\bar{\nu})$. These include models based on vector leptoquarks~\cite{Fuentes-Martin:2020hvc,Allwicher:2024ncl}, non-universal $Z'$ bosons~\cite{Athron:2023hmz} or scenarios involving combined scalars and fermions sectors~\cite{Guedes:2024vuf}. Our EFT analysis identifies the essential operator structure and flavor patterns that such UV completions must satisfy to remain consistent with the full set of flavor and electroweak constraints.

In summary, this section demonstrates that the ML framework provides
essential capabilities for this analysis: accurate emulation of the
non-Gaussian likelihood landscape with modest training requirements,
efficient generation of large representative samples for high-resolution
confidence regions and correlation studies, robust performance without
extensive hyperparameter tuning or sensitivity to initialization, and
interpretability through SHAP analysis that validates the physical
consistency of the fit. These advantages, particularly when compared to
alternative methods such as neural networks or brute-force grid
sampling, justify the central role of ML in our methodology and
demonstrate its value as a quantitative tool for precision phenomenology
in particle physics.

\section{Conclusions}
\label{sec:CONCLU}

In this work, we have performed an analysis of the anomalies in
$B$-meson decays by using an effective field theory approach and
considering the NP effects on the Wilson coefficients of the effective
Lagrangian. The new deviation observed on $B^+\to
K^+\nu\bar{\nu}$ decay at Belle II is included in the global fits.
Consequently, a new scenario (Scenario III) is considered
in the present study, with only mixing between the second and third
quark generations and no mixing in the lepton sector, being the Wilson
coefficients for the singlet and triplet four fermion effective
operators independent. Both the recent experimental measurements of
$\mathrm{BR}(B^+\rightarrow K^+\nu\bar{\nu})$ and $R_{J/\psi}$ are
included in the analysis. Besides, branching ratios associated
with $B\rightarrow D^{(*)}\tau\nu$ decays are considered. The fact that
the global fit is dominated by these classes of observables underlines
their potential as probes to NP. 

We use a ML algorithm in our analysis, by checking
the agreement between the results obtained for this procedure and both
the general shape of the $\Delta \chi^2$ distribution and the analysis
of the impact of each parameter on the global fit. In our Scenario III,
the $B\to K^{(*)}\nu\bar{\nu}$ observables constrain the parameter
$C_1$, both $B\to D^{(*)}\ell\nu$ and $B\to K^{(*)}\nu\bar{\nu}$
constrain $C_3$, and $B\to D^{(*)}\ell\nu$ observables present the most
important constraint to $\beta^q$. The results of the global fit after including a wide range of flavour and electroweak observables are displayed in Table~\ref{tab:fit_results}, and the predictions for the most relevant observables are summarized in Table~\ref{tab:comp_RKRD}. The comparison with the previous scenarios considered in~\cite{Alda:2021rgt} is included in the discussion. Our results show that Scenario III, which introduces independent parameters $C_1$ and $C_3$ without mixing in the lepton sector, provides the best fit to experimental results, achieving a pull of $6.25\,\sigma$. This scenario successfully accounts for the discrepancies observed in $R_{D^{*}}$ and $R_{J/\psi}$ measurements while maintaining consistency with the Standard Model in the $R_K$ ratios. Furthermore, it improves the predicted branching ratio for $\mathrm{BR}(B^+\rightarrow K^+\nu\bar{\nu})$ although it does not fully explain $\mathrm{BR}(B^0\rightarrow K^{*0}\nu\bar{\nu})$. It is worth noting that this tension in the neutral mode points to a limitation in the assumed flavor structure of Scenario III. Our analysis relies on left-handed SMEFT operators ($C_1, C_3$) which strictly correlate the charged and neutral channels. Reconciling the current experimental divergence between $B^{+}\to K^{+}\nu\bar{\nu}$ and $B^{0}\to K^{*0}\nu\bar{\nu}$ would likely require the inclusion of right-handed currents or more complex flavor structures beyond the minimal set considered here, as the fit is currently driven by the dominant signals in $R_{D^{(*)}}$ and the charged neutrino mode.

The results are pointing in the direction of NP that interacts mainly with the third generation of leptons, reminiscent of the $U(2)$ flavour symmetry hypothesis~\cite{Fuentes-Martin:2019mun,Faroughy:2020ina}. The situation would be clarified further with measurements of $R_{J/\psi}$ achieving a resolution similar to that of $R_{D^{(*)}}$, and with additional $\tau$ observables, for example the longitudinal polarisation in $B_c \to J/\psi \tau \nu$ or the branching ratio of $B\to K^{(*)}\tau^+\tau^-$.

We developed and applied an ML framework that emulates the highly
non-Gaussian structure of the likelihood landscape with minimal training
cost. Our results show that this method significantly improves accuracy,
efficiency, and interpretability. It enables the generation of
high-resolution confidence regions and detailed correlation
analyses. Additionally, the method remains robust to hyperparameter
choices, and offers transparent interpretability through SHAP
analysis. Together, these features establish ML as a powerful and
reliable tool for global fits  in precision particle phenomenology,
outperforming conventional approaches such as neural networks and grid
sampling. 

\section*{Acknowledgments}
This work is partially supported by Spanish MINECO/FEDER Grants
No. PGC2022-126078NB-C21 funded by MCIN/AEI/10.13039/\ 501100011033 and
“ERDF A way of making Europe” and Grant DGA-FSE Grant 2020-E21-17R
Arag\'on Government and the European Union - NextGenerationEU Recovery
and Resilience Program on {\it{Astrof\'isica y F\'isica de Altas
    Energ\'ias}} CEFCA-CAPA-ITAINNOVA. Additionally, J.A has received
funding from the Fundaci\'on Ram\'on Areces ``Beca para ampliaci\'on de
estudios en el extranjero en el campo de las Ciencias de la Vida y de la
Materia''.\\[0.3em]

"Funding information - not applicable"\\[0.5em]

\bibliographystyle{utphys.bst}
\bibliography{bibliography}

\end{document}